\documentclass[iop,twocolumn]{emulateapj}   

\usepackage{color}
\usepackage{natbib}
\usepackage{subfigure}
\usepackage{verbatim}
\usepackage{graphicx}
\usepackage{bm}              
\usepackage[amssymb]{SIunits}
\usepackage{multirow}
\usepackage{tabularx}
\usepackage{booktabs}
\usepackage{float}

\newcommand {\apgt} {\ {\raise-.5ex\hbox{$\buildrel>\over\sim$}}\ }
\newcommand {\aplt} {\ {\raise-.5ex\hbox{$\buildrel<\over\sim$}}\ }

\newcommand{\noop}[1]{}  

\newcommand{\commentx}[1]{}

\newcommand{\etal}{et al.\,}  

\newcommand{\ra}[3]   
   {\makebox[1.5em][r]{#1}\makebox[1.5em][r]{#2} \makebox[2em][r]{#3}}

\newcommand{\hms}[3]  
   {${#1}^{\mathrm{h}}{#2}^{\mathrm{m}}{#3}^{\mathrm{s}}$}

\newcommand{\hmin}[2]  
   {\ensuremath{{#1}^{\mathrm{h}}{#2}^{\mathrm{m}}}}
   
\newcommand{\hours}[1]  
   {\ensuremath{{#1}^{\mathrm{h}}}}

\newcommand{\dms}[3]  
   {\ensuremath{{#1}\degree{#2}\arcminute{#3}\arcsecond}}

\newcommand{\dm}[2]  
   {\ensuremath{{#1}\degree{#2}\arcminute}}

\newcommand{\ukcmb}  
           {\ensuremath{\micro \kelvin_\mathrm{cmb}}}

\newcommand{\uk}  
           {\ensuremath{\micro \kelvin}}

\newcommand{\fdeg} 
           {\hbox{$.\!\!^{\circ}$}}

\usepackage{color}

\slugcomment{}

\shortauthors{R.~Thornton~\etal}
\shorttitle{ACTPol Instrument}

\begin{document}


\title{THE ATACAMA COSMOLOGY TELESCOPE: THE POLARIZATION-SENSITIVE ACTPOL INSTRUMENT}

\author{
R.~J.~Thornton\altaffilmark{1,2},
P.~A.~R.~Ade\altaffilmark{3},
S.~Aiola\altaffilmark{4,5},
F.~E.~Angil\`e\altaffilmark{2},
M.~Amiri\altaffilmark{6},
J.~A.~Beall\altaffilmark{7},
D.~T.~Becker\altaffilmark{7},
H-M.~Cho\altaffilmark{8},
S.~K.~Choi\altaffilmark{9},
P.~Corlies\altaffilmark{10},
K.~P.~Coughlin\altaffilmark{11},
R.~Datta\altaffilmark{11},
M.~J.~Devlin\altaffilmark{2},
S.~R.~Dicker\altaffilmark{2},
R.~D\"{u}nner\altaffilmark{12},
J.~W.~Fowler\altaffilmark{7},
A.~E.~Fox\altaffilmark{7},
P.~A.~Gallardo\altaffilmark{10},
J.~Gao\altaffilmark{7},
E.~Grace\altaffilmark{9},
M.~Halpern\altaffilmark{6},
M.~Hasselfield\altaffilmark{13,14},
S.~W.~Henderson\altaffilmark{10},
G.~C.~Hilton\altaffilmark{7},
A.~D.~Hincks\altaffilmark{6,15},
S.~P.~Ho\altaffilmark{9},
J.~Hubmayr\altaffilmark{7},
K.~D.~Irwin\altaffilmark{8,16},
J.~Klein\altaffilmark{2},
B.~Koopman\altaffilmark{10},
Dale~Li\altaffilmark{8},
T.~Louis\altaffilmark{17},
M.~Lungu\altaffilmark{2},
L.~Maurin\altaffilmark{18},
J.~McMahon\altaffilmark{11},
C.~D.~Munson\altaffilmark{11},
S.~Naess\altaffilmark{19},
F.~Nati\altaffilmark{2},
L.~Newburgh\altaffilmark{20},
J.~Nibarger\altaffilmark{7},
M.~D.~Niemack\altaffilmark{10},
P.~Niraula\altaffilmark{21},
M.~R.~Nolta\altaffilmark{22},
L.~A.~Page\altaffilmark{9},
C.~G.~Pappas\altaffilmark{9},
A.~Schillaci\altaffilmark{9,21},
B.~L.~Schmitt\altaffilmark{2},
N.~Sehgal\altaffilmark{23},
J.~L.~ Sievers\altaffilmark{24}
S.~M.~Simon\altaffilmark{9},
S.~T.~Staggs\altaffilmark{9},
C.~Tucker\altaffilmark{3},
M.~Uehara\altaffilmark{21},
J.~van~Lanen\altaffilmark{7},
J.~T.~Ward\altaffilmark{2},
E.~J.~Wollack\altaffilmark{25}
}

\altaffiltext{1}{Department of Physics, West Chester University
of Pennsylvania, West Chester, PA 19383, USA}
\altaffiltext{2}{Department of Physics and Astronomy, University of
Pennsylvania, Philadelphia, PA 19104, USA}
\altaffiltext{3}{School of Physics and Astronomy, Cardiff University, The Parade,
Cardiff, Wales, UK CF24 3AA}
\altaffiltext{4}{Department of Physics and Astronomy, University of Pittsburgh, Pittsburgh, PA 15260, USA}
\altaffiltext{5}{Pittsburgh Particle Physics, Astrophysics, and Cosmology Center, University of Pittsburgh, Pittsburgh PA 15260, USA}
\altaffiltext{6}{Department of Physics and Astronomy, University of
British Columbia, Vancouver, BC, Canada V6T 1Z4}
\altaffiltext{7}{NIST Quantum Sensors Group, 325
Broadway Mailcode 817.03, Boulder, CO 80305, USA}
\altaffiltext{8}{SLAC National Accelerator Laboratory, 2575 Sand Hill Road, Menlo Park, CA 94025}
\altaffiltext{9}{Joseph Henry Laboratories of Physics, Jadwin Hall,
Princeton University, Princeton, NJ  08544, USA}
\altaffiltext{10}{Department of Physics, Cornell University,
Ithaca, NY  14853, USA}
\altaffiltext{11}{Department of Physics, University of Michigan
Ann Arbor, MI  48109, USA}
\altaffiltext{12}{Instituto de Astrof\'isica and Centro de 
Astro-Ingenier\'ia, Facultad de F\'isica, Pontificia Universidad Cat\'olica de Chile, Av. Vicuña Mackenna 4860, 7820436 Macul, Santiago, Chile}
\altaffiltext{13}{Department of Astronomy and Astrophysics, The Pennsylvania State University, University Park, PA 
16802, USA}
\altaffiltext{14}{Institute for Gravitation and the Cosmos, The Pennsylvania State University,
   University Park, PA 16802, USA}
\altaffiltext{15}{Pontificia Universit\'a Gregoriana, Piazza della Pilotta 4, 00187 Roma, Italy}
\altaffiltext{16}{Department of Physics, Stanford University,
Stanford, CA  94305, USA}
\altaffiltext{17}{UPMC Univ Paris 06, UMR7095, Institut d'Astrophysique de Paris, F-75014, Paris, France}
\altaffiltext{18}{Instituto deAstrof\'isica, Facultad de F\'isica, Pontificia Universidad Cat\'olica de Chile, Av. Vicuña Mackenna 4860, 7820436 Macul, Santiago, Chile}
\altaffiltext{19}{Sub-Department of Astrophysics, University of Oxford, Keble Road, Oxford, UK OX1 3RH}
\altaffiltext{20}{Dunlap Institute for Astronomy and Astrophysics,
University of Toronto, Toronto, Ontario, Canada M5S 3H14}
\altaffiltext{21}{Sociedad Radiosky Asesorıas de Ingenier\'ia Limitada Lincoy\'an 54, Depto 805 Concepci\'on, Chile}
\altaffiltext{22}{Canadian Institute for Theoretical Astrophysics, University of
Toronto, Toronto, ON, Canada M5S 3H8}
\altaffiltext{23}{Physics and Astronomy Department, Stony Brook University, Stony Brook, NY 11794-3800, USA}
\altaffiltext{24}{Astrophysics and Cosmology Research Unit, School of Chemistry and Physics, University of KwaZulu-Natal, Durban, South Africa
National Institute for Theoretical Physics, KwaZulu-Natal, South Africa}
\altaffiltext{25}{NASA/Goddard Space Flight Center, Greenbelt, MD  20771, USA}

\begin{abstract}
The Atacama Cosmology Telescope (ACT) is designed to make high angular resolution measurements of anisotropies in the Cosmic Microwave Background (CMB) at millimeter wavelengths. We describe ACTPol, an upgraded receiver for ACT, which uses feedhorn-coupled, polarization-sensitive detector arrays, a 3$^{\circ}$ field of view, 100\,mK cryogenics with continuous cooling, and meta material anti-reflection coatings. ACTPol comprises three arrays with separate cryogenic optics: two arrays at a central frequency of 148\,GHz and one array operating simultaneously at both 97\,GHz and 148\,GHz. The combined instrument sensitivity, angular resolution, and sky coverage are optimized for measuring angular power spectra, clusters via the thermal Sunyaev-Zel'dovich and kinetic Sunyaev-Zel'dovich signals, and CMB  lensing due to large scale structure. The receiver was commissioned with its first 148\,GHz array in 2013,  observed with both 148\,GHz arrays in 2014, and has recently completed its first full season of operations with the full suite of three arrays. This paper provides an overview of the design and initial performance of the receiver and related systems.  
\end{abstract}
\keywords{Microwave Telescopes, CMB Observations}


\section{Introduction}
\label{sec_introduction}

Measurements of Cosmic Microwave Background (CMB) temperature anisotropies at scales from one arcminute to many degrees have placed precise constraints on the $\Lambda$CDM cosmological model. Recent results come from, e.g., the WMAP collaboration \citep{Hinshaw2013}, Planck \citep{Planck2015}, the Atacama Cosmology Telescope (ACT; \citealp{Sievers2013}), and the South Pole Telescope (\citealp{Hou2014}). ACT was commissioned in 2008 with its first instrument, the Millimeter Bolometer Array Camera (MBAC; \citealp{swetz2011}). MBAC had three independent sets of optics at frequencies of 148\,GHz, 218\,GHz, and 277\,GHz, and was not sensitive to polarization.  In this paper we describe the polarization-sensitive second generation camera, ACTPol.

Measurements of CMB polarization provide independent constraints on cosmological parameters, and have lower foreground power at small scales because dusty sources are relatively unpolarized. Polarization can be separated into ``curl-free" E-modes and ``divergence-free" B-modes (\citealp{seljak1998}; \citealp{kamionkowki1998}). For $\ell \gtrsim 20$, E-modes arise primarily from density perturbations in the early Universe.  B-modes at $\ell \gtrsim$ 50 arise from the gravitational lensing of the CMB by large scale structure. At $\ell \lesssim 120$, gravitational waves generated during an inflationary epoch would also generate B-modes.  

To mine the rich science potential of the polarized CMB, we have built a new, polarization-sensitive instrument for ACT.  The ACTPol receiver \citep{Niemack2010} contains three different sets of of mm-wave optics, detector arrays, and associated multiplexing readouts.  Polarization-sensitive arrays 1 and 2 (hereafter``PA1" and ``PA2") operate\footnote[1]{Operating frequencies are based on effective CMB band centers (Table \ref{tab:bandcenters}), where the 148 GHz nominal frequency is based on an average across all three arrays.  The 146\,GHz central frequency for PA1 in \cite{Naess2014} was based on pre-deployment lab measurements made with a reduced Lyot stop.} at 148\,GHz, and the third polarization array (``PA3")  operates at both 97\,GHz and 148\,GHz (\citealp{mcmahon2012}; \citealp{datta/etal:2014}).  The 148\,GHz principal frequency is driven by three considerations: its location in an atmospheric window between oxygen and water lines  (Section~\ref{sec_filters}), sensitivity to the Sunyaev-Zel'dovich (SZ) effect and the CMB, and angular resolution.  At higher frequencies, resolution is better but atmospheric opacity increases; at lower frequencies, resolution is poorer but the atmosphere is more transparent. Each set of optics has a field of view (FOV) that spans approximately 1$^\circ$  on the sky and illuminates an array of corrugated feedhorns which, in turn, direct light to pairs of transition edge sensor (TES) bolometers (one for each orthogonal polarization). Each feedhorn in the PA3 array couples to four bolometers, with an orthogonal pair at 148\,GHz and another pair at 97\,GHz.  Combined, there are 1279 feedhorns and 3068 detectors in the three arrays.

This paper details the design and initial performance of ACTPol. First light with PA1 was achieved in June 2013. Both PA1 and PA2 operated in 2014. First light with all three arrays installed was in February 2015.  Section\,\ref{sec_site} provides a brief review of the ACT site and telescope.  Our choice of bands and observing strategy are described in Section~\ref{sec_sky}. The details of the cryogenic instrument are discussed in Sections~\ref{sec_coldopt} (optics), \ref{sec_arrays} (cryostat), and \ref{sec_cryostat} (detectors).    Section~\ref{sec_control} outlines the relevant data acquisition systems, and Section 8 presents on-sky performance.


\section{Observing Site and Telescope Design}
\label{sec_site}

The ACT site is located at an altitude of 5190 m on the slopes of Cerro Toco in the Atacama Desert of northern Chile.  At a latitude of 23$^\circ$~S, the telescope has access to over half of the sky. The high altitude and low precipitable water vapor (PWV) at this location provide excellent millimeter and submillimeter atmospheric transparency, with optimum observing between April and December, when the weather is the driest.    Further discussion of the site characterization during the ACT observing seasons, including PWV, observing efficiency, and atmospheric fluctuations, can be found in \citet{Dunner:2009}.  The  advantages of this location have attracted a number of other millimeter experiments, including the Atacama B-mode Search \citep{essinger-hileman:2011}, the POLARBEAR instrument \citep{Kermish2012}, and the Cosmology Large Angular Scale Surveyor \citep{essinger-hileman:2014}.

\begin{figure}
\includegraphics[width=3.4in]{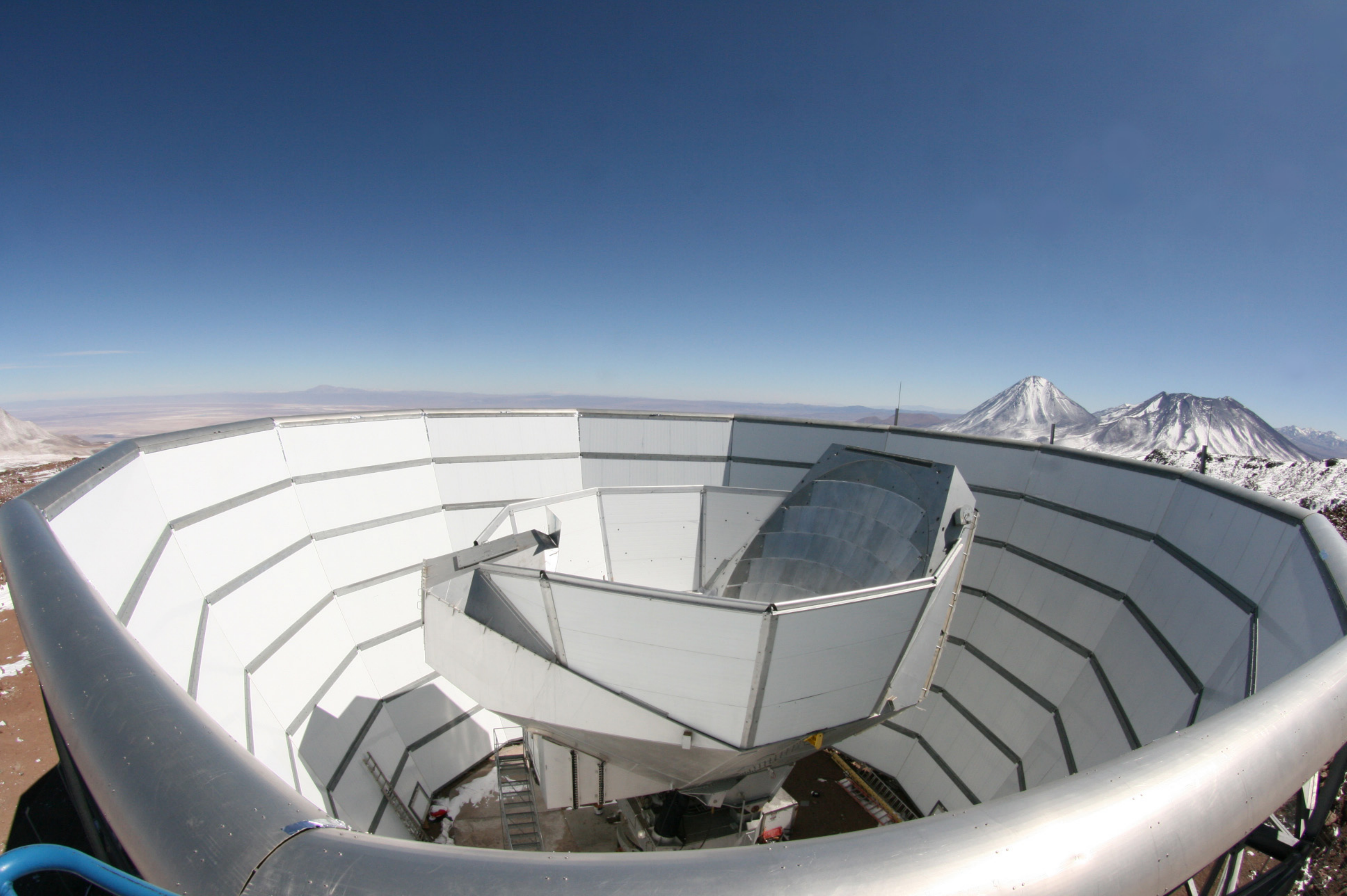}
 \caption{Photograph of ACT inside its stationary ground screen, which shields the telescope from ground emission. The secondary mirror is barely visible behind the inner, co-moving ground screen.  For scale, from the ground to the top of the telescope is 12\,m.}
\label{fig:telescopepic}
\end{figure}

\begin{figure*}[!t]
\begin{center}
\includegraphics[width=6in]{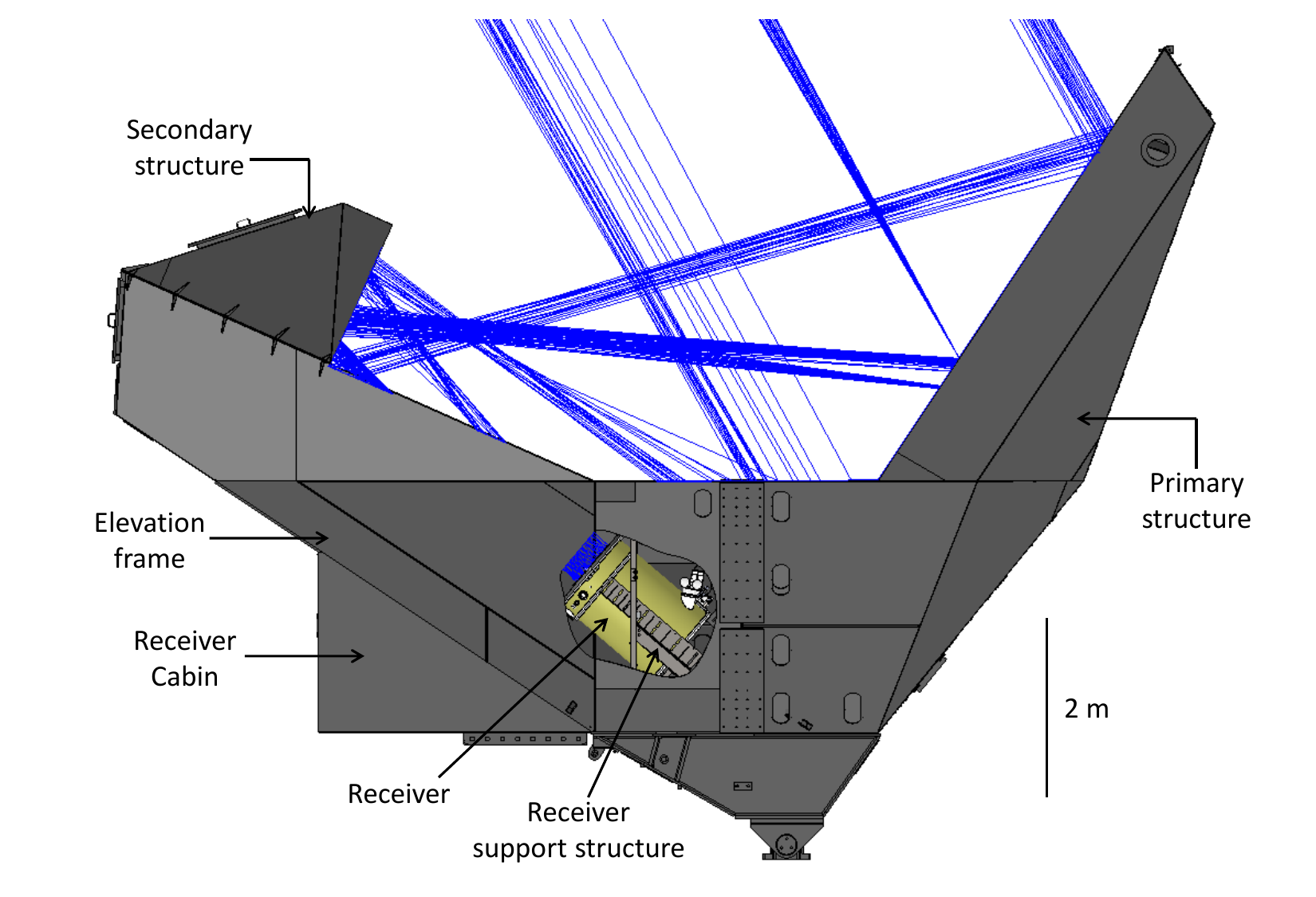}
\end{center} 
\caption{Ray trace of ACT's primary and secondary mirrors up to the entrance of the receiver.  The major components of the telescope upper structure are shown, except for the inner ground screen and part of the receiver cabin wall, which have been removed for clarity.  The telescope is shown in its service position (where the receiver cabin floor is level), corresponding to a viewing elevation of 60$^{\circ}$.}
\label{fig:telescoperaytrace}
\end{figure*}

The telescope is a numerically optimized off-axis Gregorian design. The size of the 6-m primary was dictated by the requirement for arcminute resolution.  The primary's 5.2-m focal length allows for a compact arrangement and the ability for fast scanning. The major components of the telescope are shown in Figures~\ref{fig:telescopepic} and \ref{fig:telescoperaytrace}. The primary and secondary mirrors are composed of 71 and 11 adjustable aluminum panels, respectively. The mirrors are surrounded by an inner ground screen that moves with the telescope during scanning, and an outer, stationary ground screen. A climate-controlled receiver cabin is situated underneath the primary and secondary mirrors. A more thorough discussion of the telescope optical design can be found in  \citet{swetz2011} and \citet{fowler2007}.


\section{Sky Coverage and Scan Strategy}
\label{sec_sky}

Two general survey strategies, ``wide'' and ``deep," have been implemented with ACTPol.  The deep observing strategy targets four $\sim70$ deg$^2$ regions that overlap with other rich multi-wavelength surveys.  These deep fields are near the equator and spaced a few hours apart, enabling rising and setting observations of the different regions to be interspersed.  Observations of the deep fields were made in 2013 with the PA1 148\,GHz array; a summary of these along with first results are presented in \citet{Naess2014}.  The wide observing strategies implemented in 2014 and 2015 cover a few thousand square degrees and primarily overlap with the SDSS Baryon Oscillation Spectroscopic Survey and other equatorial observations.  Southern regions with lower galactic foreground emission have also been observed, especially during the daytime when the Sun traverses the equatorial fields.  

To separate the CMB signal from drifts in the detectors and atmosphere, the entire upper telescope structure, including primary, secondary, inner ground screen, and receiver, is scanned in azimuth at a fixed elevation to minimize the effects of changing airmass. Most scans are performed both east (rising) and west (setting) so that regions can be observed at a range of parallactic angles, thus allowing separation of instrumental and celestial polarization, and improving the dynamic range of the maps by providing cross-linking.

The CMB fields are observed by scanning at 1.5~deg/sec in azimuth with 0.4\,sec turnarounds. ACT is typically operated in the elevation range of 40$^\circ$ to 55$^\circ$ for science observations, which gives it access to a significant fraction of the sky. The duration of the scan generally takes between 10 and 20 sec, depending on elevation and the target field.  When the elevation is changed, the detector bias is modulated to recalibrate and check for any changes in the time constants (Section~5) due to the change in sky load.


\section{Cryogenic Optics}
\label{sec_coldopt}
\subsection{Optics Overview}

\begin{figure}[!b]
\begin{center}
\includegraphics[width=3.35in]{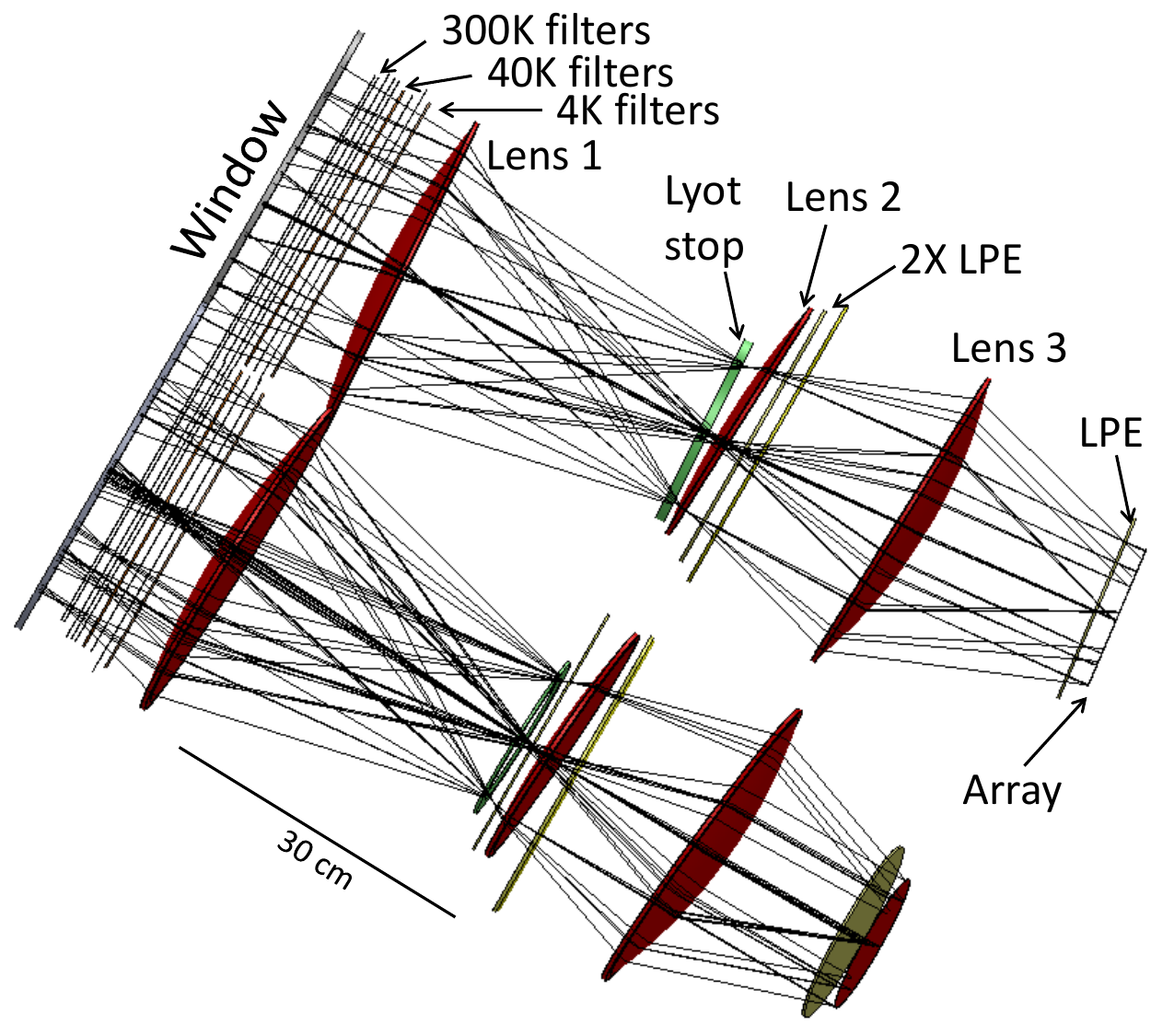}
 \end{center}
\caption{Ray trace of the cold optics.  The upper trace shows the PA3 (multichroic) optical path and the lower trace shows the PA1 path.  The PA2 optical path is a mirror image to that of PA1 and has been removed for clarity. The constituent elements are described in the text.}
\label{fig:cryo_raytrace}
\end{figure}

Three independent cryogenic optics tubes are used to reimage the Gregorian focus of the telescope  onto the detector arrays. These refractive optics are designed to maximize the optical throughput and instrument sensitivity.     Each optics tube uses three anti-reflection (AR) coated silicon lenses to reimage a $\sim$ 1$^\circ$ diameter portion of the Gregorian focus.   The primary elements of each set of camera optics are a cryostat window and IR blocking filters \citep{Tucker2006} at ambient temperature; followed by a combination of blocking filters and low pass edge (LPE) filters \citep{ade2006} at 40\,K; the first lens and accompanying filters at 4\,K;  the Lyot stop, two more lenses, and additional low pass filters all at 1\,K; and the final LPE filter and array package at 100\,mK. A ray trace of the cold optics is shown in Figure~\ref{fig:cryo_raytrace}.    

The size of each optics tube is limited by both the size of the cryostat (which had to fit in the existing receiver cabin) as well as the the maximum diameter of the low-pass edge filters (Section~\ref{sec_filters}).  To minimize the size of the entrance optics,  the receiver is positioned such that the Gregorian focus is located between the receiver window and first lens.   The Gregorian focus is not telecentric, which is a requirement for a large, flat feedhorn-coupled detector array (see \citealt{hanany2013}).  To achieve a telecentric design, small offsets and tilts were incorporated into the three lenses.  The final design is diffraction-limited across each focal plane.

\subsection{Lenses}
\label{sec_lenses}
\begin{figure}[!h]
\begin{center}
\includegraphics[width=3.25in]{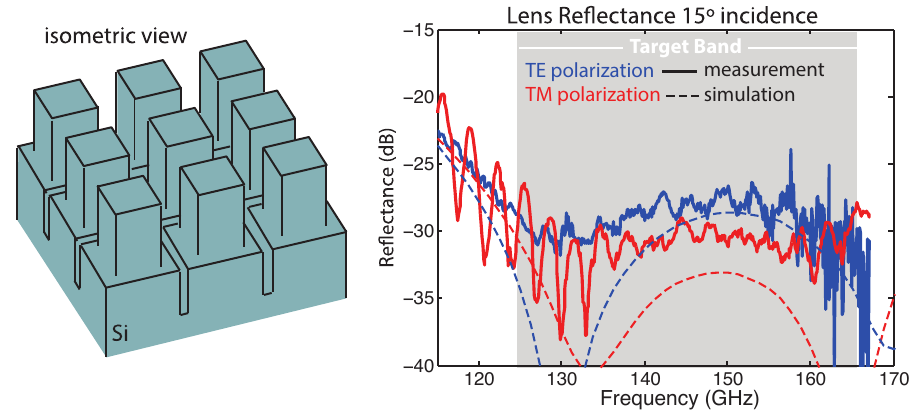}
 \end{center}
\caption{(Left) Isometric view of a two-layer antireflection coating showing how the material removal process creates ``pillars" on the lens surface. For scale, the pillar pitch  is 450\,$\mu$m. (Right) Comparison between simulated and measured reflectance of a two-layer coating on one side of a flat silicon sample.}
 \label{fig:ARcoating}
\end{figure}

Silicon was chosen for the lens material due to its high thermal conductivity, high index of refraction (n\,=\,3.4), and low loss at our wavelengths.  The high index of refraction necessitates the use of AR coatings.  ACT previously used Cirlex coatings \citep{lau2006a}, but they incurred an estimated 15\% net efficiency reduction \citep{swetz2011}.   For ACTPol, we created ``meta material" AR coatings produced by removing some of the silicon to controlled depths from the surfaces of each lens at sub-wavelength scales using a custom three-axis silicon dicing saw, creating layers of square pillars.  The resulting coating has a coefficient of thermal expansion matching that of the rest of the lens.  Lenses based on a two-layer design (Figure~\ref{fig:ARcoating}a) are used in both the PA1 and PA2 optics.  Simulations showed that the resulting coating has low-reflection ($<$\,1\%) for angles of incidence up to  30$^\circ$ with low cross-polarization \citep{datta/etal:2013}. Figure~\ref{fig:ARcoating}b shows measurements verifying the simulated performance.  The meta material method was extended to a three-layer pillar design for the wider bandwidth of PA3.  The measured performance agrees with predictions, and both PA1/2 and PA3 are consistent with sub-percent level reflections \citep{datta/etal:2015}.  

\subsection{Filters and Bandpasses}
\label{sec_filters}

Each optics tube has its own circular, 6.4\,mm-thick window made of ultra-high molecular weight polyethlene with an expanded Teflon AR coating.  The PA1 and PA2 windows are 32\,cm in diameter and the PA3 window is 34\,cm in diameter.  Although thinner windows would have reduced in-band loading, the increased deformation would have interfered with the blocking filters immediately behind them (Figure~\ref{fig:cryo_raytrace}).  There are IR blocking filters \citep{Tucker2006} at 300\,K, 40\,K, and 4\,K that reflect high-frequency out-of-band radiation to reduce the optical load, in particular, on the poorly thermally conducting LPE filters.   These LPE capacitive mesh filters \citep{ade2006} are used at 40K, 4K, 1K, and 100 mK to limit loading on successive stages.  Filter sets with a range of cutoff frequencies allow suppression of out-of-band leaks from individual filters.  For PA1/2, these LPE filters are also used to define the upper edge of the band. The lower edge of the PA1/2 band is set by a waveguide cutoff at the end of each feedhorn (Section~\ref{sec_feedhorns}).  The PA3 bands are set by on-chip filters (Section~\ref{sec_detectors}).

\begin{figure}[h]
\begin{center}
\includegraphics[width=3.7in]{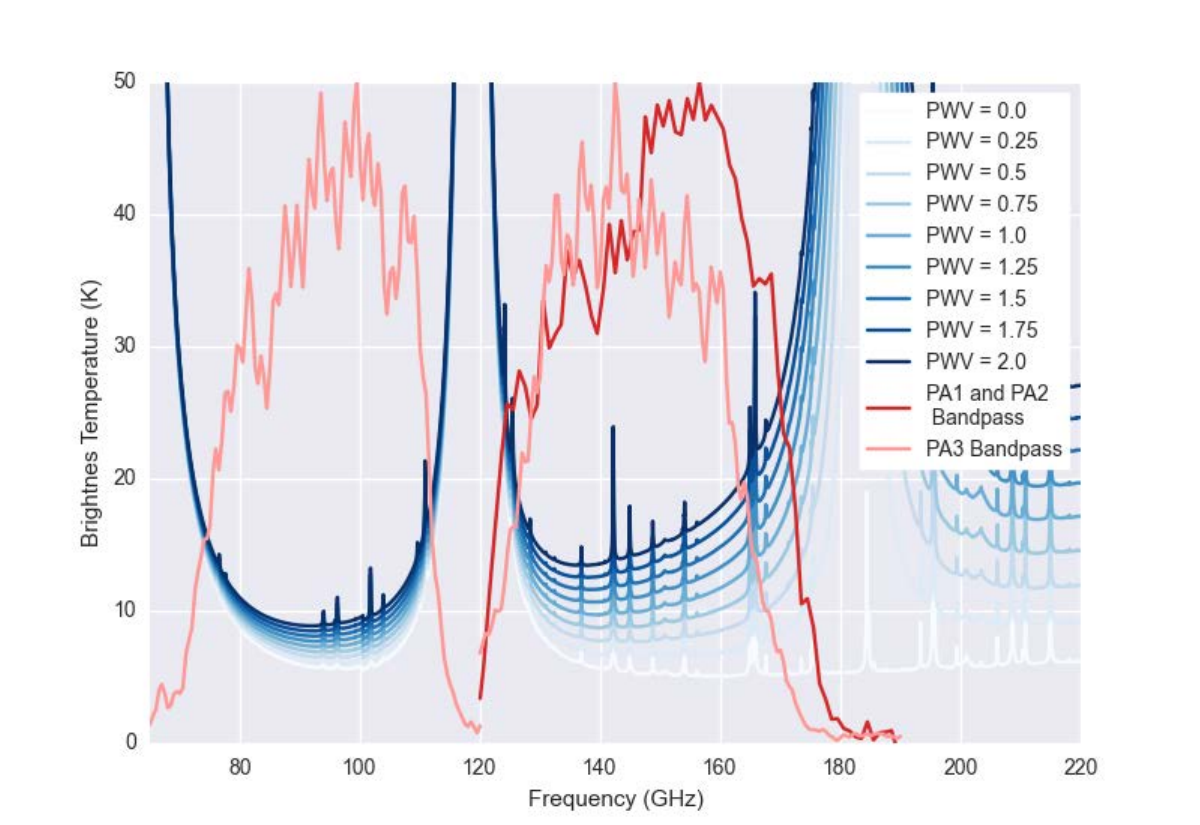}
\end{center}
\caption{Plot of the ACTPol bandpasses for all three arrays (shown with arbitrary scaling) superimposed on the atmospheric brightness displayed for a range of PWVs.  The bandpasses shown are an average of 21 detectors for PA1, 84 detectors for PA2, 18 detectors for PA3 97\,GHz, and 28 detectors for PA3 148\,GHz. The predominant atmospheric features are the 60\,GHz and 117\,GHz oxygen lines and the 183\,GHz water line.  The loading due to water increases with the atmospheric water vapor content while the oxygen emission remains relatively constant. Because of its proximity to the 183\,GHz line, the 148\,GHz band is significantly more sensitive to weather variation than the 97\,GHz band. The atmospheric transmission was generated with the ALMA Atmospheric Transmission Modeling (AATM) code, which is a repackaging of the ATM code described in \citet{pardo:2001}.\\}
\label{fig:bandpass}
\end{figure}

Each filter's frequency response at room temperature was measured with a Fourier transform spectrometer (FTS).   After the filters were installed in the receiver, additional FTS measurements were made on the fully cooled system before shipping to the site, as were tests with thick-grill filters (e.g., \citealp{timusk1981}) to check for high frequency leaks.  A final set of FTS measurements was performed after the receiver was installed on the telescope so that bandpasses could be taken in the same configuration as science data acquisition, and are shown in Figure~\ref{fig:bandpass}.  

\begin{table*}[ht]
\centering
\begin{tabular}{ccccc}
\midrule \midrule
\bf{Array} & \bf{PA1 } & \bf{PA2} & \bf{PA3 (lower)} & \bf{PA3 (upper)} \\
\midrule
\em{Effective bandwidth} & 51 $\pm$ 1.5 & 51 $\pm$ 1.5 & 38 $\pm$ 1.5 & 41 $\pm$ 1.5 \\
\midrule \midrule
\em{Effective band centers:} & \multicolumn{4}{c}{\em compact sources} \\
\midrule
Synchrotron & 143.9 $\pm$ 2.4 & 144.2 $\pm$ 2.4 & 91.0 $\pm$ 1.7 & 143.1 $\pm$ 2.4 \\
Free-free & 144.7 $\pm$ 2.4  & 145.0 $\pm$ 2.4 & 91.8 $\pm$ 1.7  & 143.7 $\pm$ 2.4 \\
Rayleigh-Jeans  & 147.7 $\pm$ 2.4& 148.0 $\pm$ 2.4& 94.6 $\pm$ 1.7  & 145.6 $\pm$ 2.4 \\
Dusty source  & 150.0 $\pm$ 2.4 & 150.3 $\pm$ 2.4& 96.7 $\pm$ 1.7 & 147.2 $\pm$ 2.4 \\
\midrule \midrule
\em{Effective band centers:} & \multicolumn{4}{c}{\em diffuse sources} \\
\midrule
Synchrotron & 145.3 $\pm$ 2.4 & 145.6 $\pm$ 2.4 & 92.4 $\pm$ 1.7 & 144.1 $\pm$ 2.4 \\
Free-free & 146.1 $\pm$ 2.4  & 146.4 $\pm$ 2.4& 93.2 $\pm$ 1.7 & 144.6 $\pm$ 2.4 \\
Rayleigh-Jeans  & 149.1 $\pm$ 2.4& 149.4 $\pm$ 2.4& 95.9 $\pm$ 1.7  & 146.6 $\pm$ 2.4\\
Dusty source & 151.4 $\pm$ 2.4& 151.7 $\pm$ 2.4& 97.9 $\pm$ 1.7  & 148.1 $\pm$ 2.4 \\
CMB & 148.9 $\pm$ 2.4 & 149.1 $\pm$ 2.4 & 97.1 $\pm$ 1.7  & 146.6 $\pm$ 2.4 \\
SZ effect & 145.0 $\pm$ 2.4 & 145.3 $\pm$ 2.4& 94.1 $\pm$ 1.7 & 143.7 $\pm$ 2.4 \\
\midrule \midrule
{\em Conversion factors} & & & & \\
\midrule
$\delta$T$_{\rm CMB}$/$\delta$T$_{\rm RJ}$ & 1.72 $\pm$ 0.03 & 1.72 $\pm$ 0.03 & 1.27 $\pm$ 0.02 & 1.69 $\pm$ 0.03 \\
$\delta$W/$\delta$T$_{\rm RJ}$ (pW/K) & 1.41 $\pm$ 0.04 & 1.41 $\pm$ 0.04 & 1.05 $\pm$ 0.04 & 1.13 $\pm$ 0.04 \\
$\Gamma_{\rm RJ} (\micro \rm{W/Jy})$  & 7640 $\pm$ 300 & 8160 $\pm$ 290 & 5740 $\pm$ 300 & 6160 $\pm$ 300 \\
$\Gamma_{\rm CMB} (\micro \rm{W/Jy})$  & 13160 $\pm$ 510 & 14070 $\pm$ 490 & 9720 $\pm$ 520 & 7820 $\pm$ 380 \\
\midrule \midrule
\em{Beam solid angles (nsr)} & 192 $\pm$ 4 & 179 $\pm$ 3 & 264 $\pm$ 11 & 560 $\pm$ 19 \\
\midrule
\end{tabular}
\caption{Effective central frequencies for broadband compact and diffuse sources.  The values are obtained by averaging over individual detectors and based on the measured response of the receiver on the telescope.  Assumed spectral indices: synchrotron emission ($\alpha=-0.7$), free-free emission ($\alpha=-0.1$), Rayleigh Jeans emission ($\alpha=2.0$), dusty source emission ($\alpha=3.7$).  The uncertainties are obtained from a combination of the variance of the measurements and an estimate of systematic error.}
\label{tab:bandcenters}
\end{table*}

Using each array's measured bandpass, we follow the method of \citet{page2003} for calculating the effective central frequency for broadband sources, as well as the CMB and SZ effect.  The effect of the varying source spectra on the band center is to shift it slightly.  The results are given in Table\,\ref{tab:bandcenters}.


\section{Arrays}
\label{sec_arrays}

ACTPol has three superconducting focal plane arrays, each consisting of silicon micro-machined feedhorns that direct light to matched, photon-limited bolometric detector arrays \citep{yoon2009}.   The feedhorns, detectors, and SQUID multiplexing components are fabricated at NIST. PA1 and PA2 each has 512 feeds, which couple to 1024 bolometers (one per linear polarization) and operate at 148\,GHz; PA3 has 255 feeds that couple to 1020 bolometers (one per linear polarization per frequency) and operate simultaneously at 97 and 148\,GHz.  Each array is approximately 15\,cm in diameter and is located behind a 100\,mK LPE filter (Figure~\ref{fig:cryo_raytrace}).

\subsection{Feedhorns}
\label{sec_feedhorns}

\begin{figure}[h!]
\centering
\mbox{
\subfigure{\includegraphics[height=2.2in]{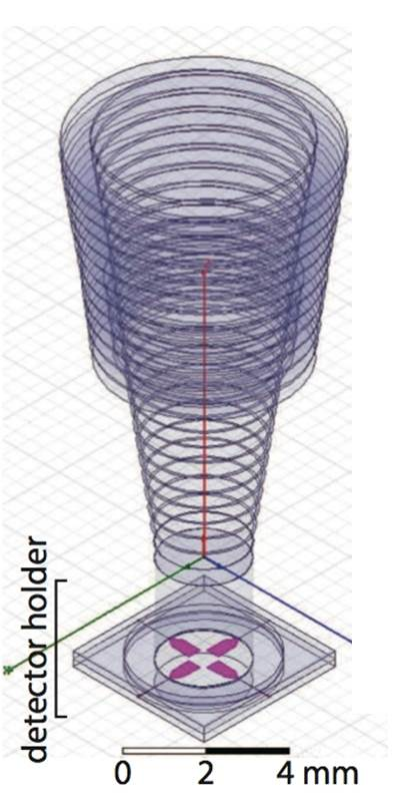}}\hspace{1cm}
\subfigure{\includegraphics[height=2.0in]{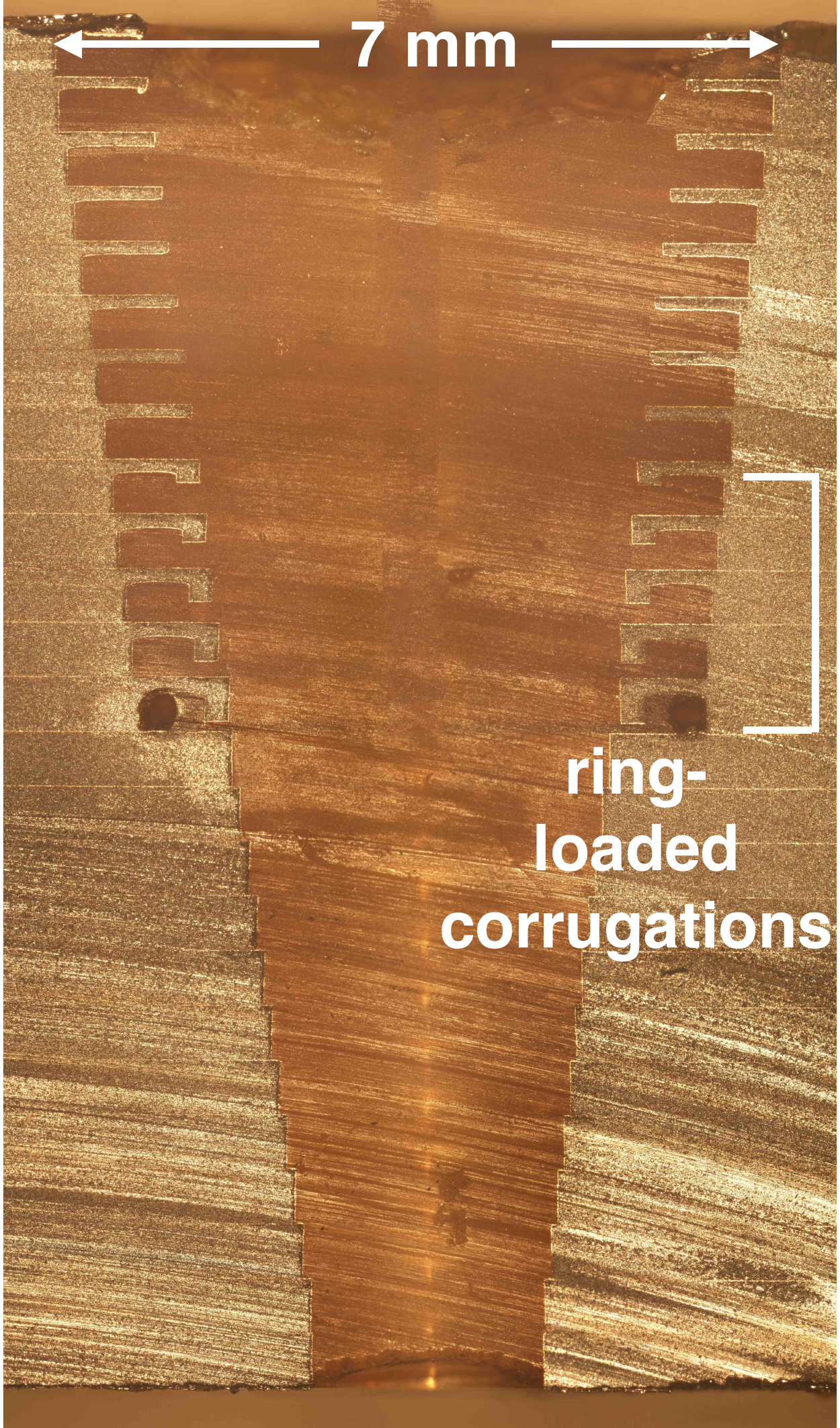}}
}
\caption{(Left) Design of a single horn-coupled multichroic polarimeter, consisting of a broad-band ring-loaded corrugated feedhorn and a planar detector array.  A broad-band Ortho-Mode Transducer (OMT), shown in magenta, separates the incoming radiation according to linear polarization.  (Right) Photograph of a cross-section of a PA3 test feedhorn.  There are a total of 25 gold-plated silicon wafers, five of which are ring-loaded to form a broad-band impedance matching transition between the corrugated input waveguide and the round output guide.}
 \label{fig:feedhorn-to-tes}
\end{figure}

The feedhorns impedance-match radiation from free space to the detector wafers and their dimensions are optimized for a suitable bandwidth (\citealp{Britton2010}; \citealp{nibarger2012}).  The feedhorn arrays are assembled from stacks of silicon wafers with micro-machined circular apertures that correspond to individual corrugations (Figure~\ref{fig:feedhorn-to-tes}). The individual silicon wafer platelets are etched, RF sputter coated with a Ti/Cu layer on both sides, stacked, aligned, and finally copper and gold electroplated to form a close-packed feedhorn array. This approach preserves the advantages of corrugated feedhorns such as low sidelobes, low cross-polarization, and wide-band performance while reducing the difficulty of building such a large array using traditional techniques like direct machining or electroforming individual metal feeds.  While similar arrays can be made from metal platelets, difficulties associated with weight, thermal mass, and differential thermal contraction are avoided with the silicon platelet array concept.   Furthermore, tolerances achievable with optical lithography and silicon micro-machining make for extremely good array uniformity.  The feedhorn profiles for all three arrays contain sections to define the low-frequency cutoffs of the detector bandpasses.  The wider band of  PA3 motivated a section of ring-loaded platelets (\citealp{Takeichi1971}) for that array (see Figure~\ref{fig:feedhorn-to-tes} and \citealt{mcmahon2012}).

\begin{figure*}[t]
\includegraphics[width=0.98\textwidth]{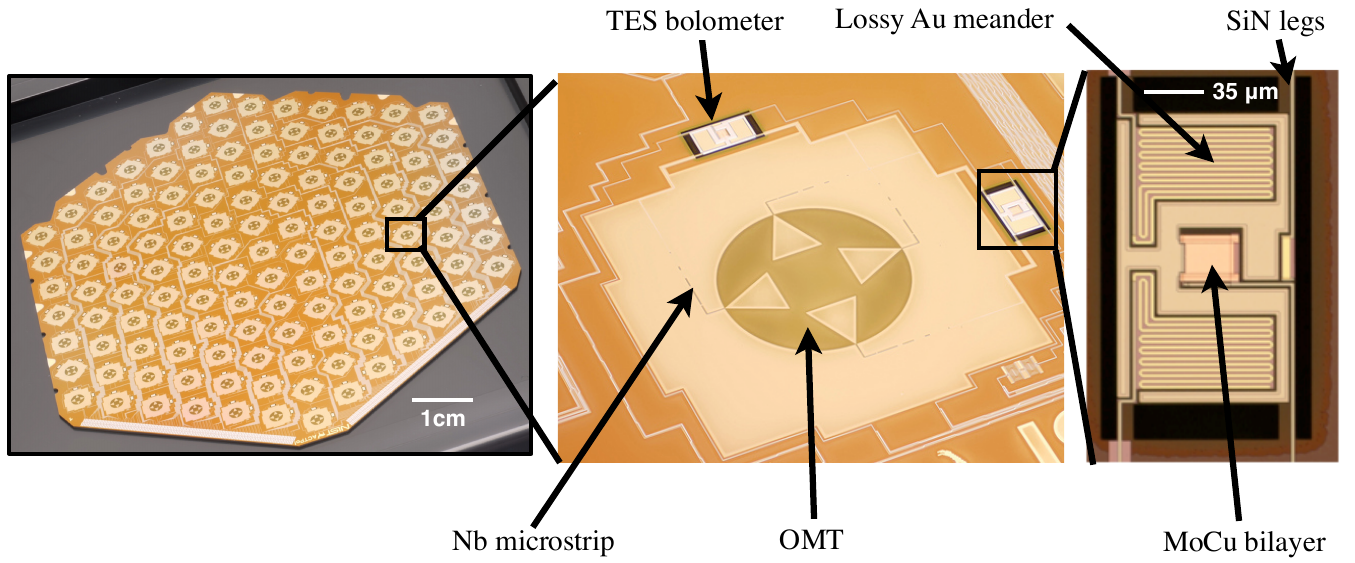}
\caption{(Left) An ACTPol 148\,GHz hex wafer containing \mbox{127 pixels} and \mbox{254 TESes}.  Each array consists of three hex wafers and three semihex wafers for a total of \mbox{522 pixels}.  (Middle) A single ACTPol pixel.  The incident radiation couples to each detector via the central OMT antenna (the different fin shape than that in Figure\,\ref{fig:feedhorn-to-tes} is a characteristic of the single-band versus multichroic design).  Niobium microstrip lines carry the power to one of the two bolometer islands on the edges.  (Right) An ACTPol TES bolometer.  The power is deposited on the island as heat through the lossy Au meander.  The island is thermally isolated from the bath, connected only through four SiN legs which also carry the TES bias and signal lines.  The electrical power dissipated in the voltage biased MoCu superconducting ``bilayer" monitors the power deposited in the lossy Au meander.}
\label{fig:wafer}
\end{figure*}

\subsection{Detectors}
\label{sec_detectors}

Behind the feedhorns, radiation is coupled onto the polarimeters via planar orthomode transducers (OMT).  The detectors are fabricated on monolithic three-inch silicon wafers in two different varieties: a hexagonal (``hex") wafer and a semi-hexagonal (``semihex") wafer.  Each hex wafer has 127 pixels in the 148\,GHz design and 61 pixels in the dichroic design, while the semihexes have 47 and 24 pixels, respectively.   Each hex and semihex section is actually composed of four silicon wafers: a top, waveguide interface wafer, an OMT and detector wafer, and a two-piece quarter-wave backshort wafer.  A full ACTPol array is assembled from three hex wafers and three semihex wafers.

Changes in radiation power are detected using superconducting TES bolometers (\citealp{Lee1996}; \citealp{irwin2005}).  When a TES is appropriately voltage biased, negative electrothermal feedback maintains it at the transition temperature, $T_c$, under a wide range of observing conditions.  Its steep resistance-vs-temperature curve transduces temperature fluctuations into current fluctuations, which are read out by the use of superconducting quantum interference device (SQUID) amplifiers using a time-domain multiplexing architecture \citep{reintsema2003} and room temperature electronics \citep{battistelli2008}.  

\begin{figure*}[t]
\centering
\mbox{
\subfigure{\includegraphics[height=2.2in]{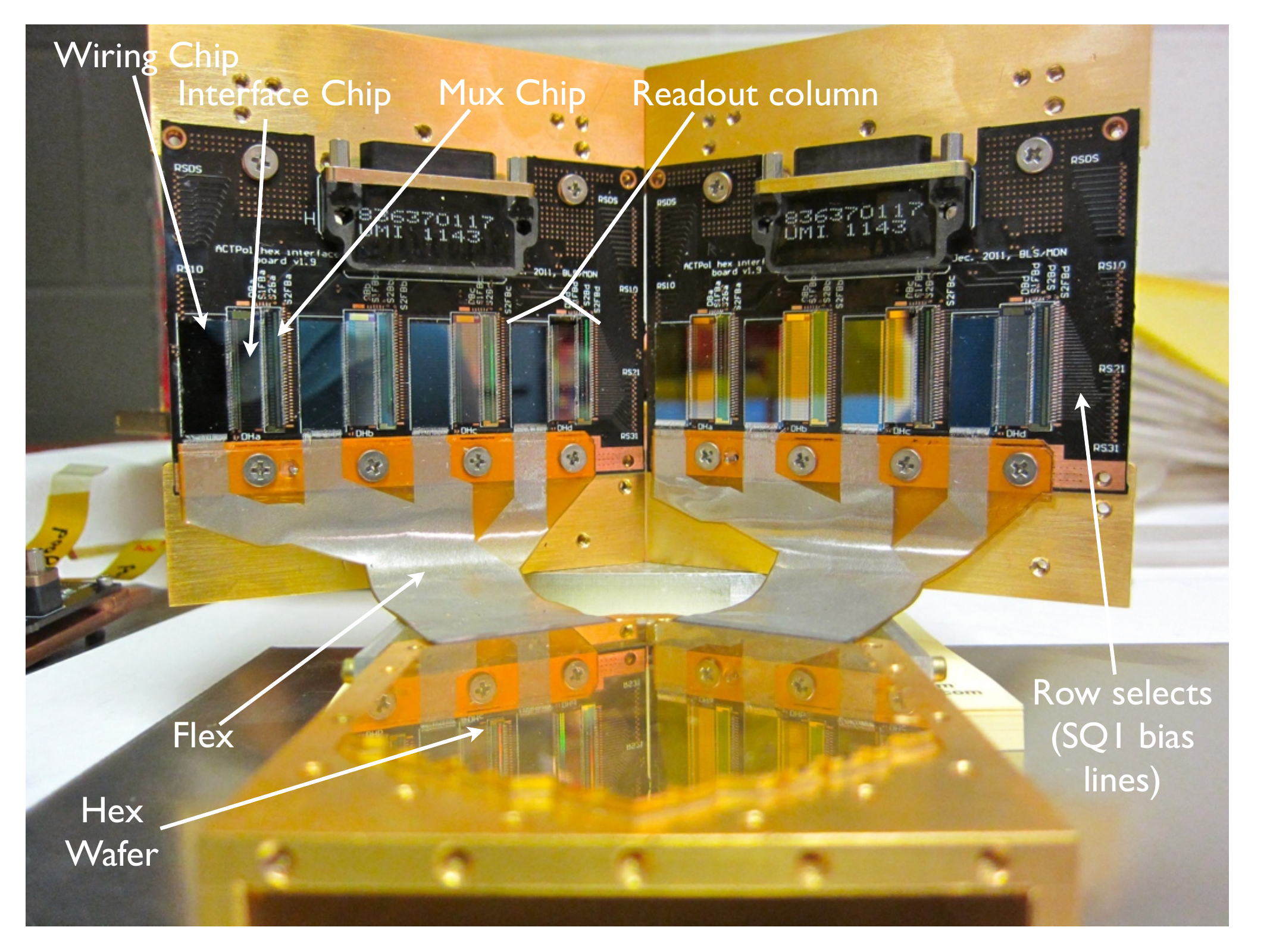}}\quad
\subfigure{\includegraphics[height=2.2in]{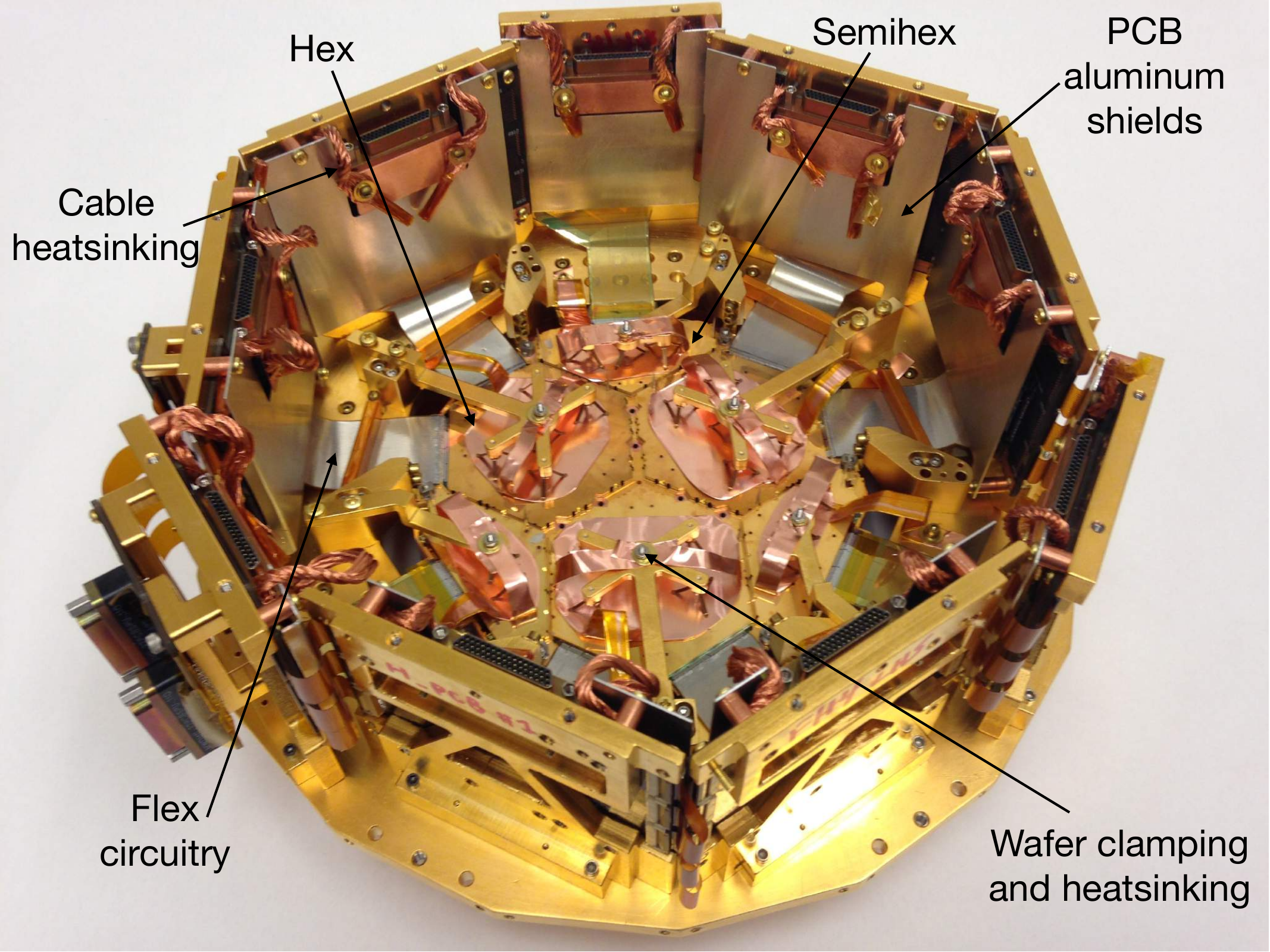}}
}
\caption{(Left) An assembled hex wafer viewed edge-on.   A hex is read out using two PCBs, each serving four readout columns. The circuitry of each column includes a mux chip, an interface chip, and a wiring chip. The wiring chip provides superconducting 90$^{\circ}$ bend routing from the flex wire bonds to the interface chip wire bonds. Connections to the PCB for the SQ1 bias and feedback lines, SQ2 bias and feedback lines, and detector bias lines are made via aluminum wire-bonds.  The detector bias lines are carried from the PCB to the wafer via the folded flex which is wire-bonded at either end. (Right) A fully assembled ACTPol array (PA2).  The hex and semihex wafers are sitting on the feedhorn array in the center with the readout PCBs arranged vertically around the edge.  The feedhorn apertures are pointed down in this photo.}
\label{fig:array}
\end{figure*}

Photographs of a hex wafer, an ACTPol pixel, and a TES bolometer for the 148\,GHz band are shown in \mbox{Figure~\ref{fig:wafer}}. Opposing pairs of OMT antenna probes separate the orthogonal polarization signals.  The signals travel through a coplanar waveguide to microstrip transition, which is impedance-matched to reduce loss, to Nb microstrip lines, where they exit. On a PA3 dichroic pixel, diplexers comprised of two separate five pole resonant stub band-pass filters separate the radiation into \mbox{75-110 GHz} and \mbox{125-170 GHz} pass-bands. The signals from opposite OMT probes within a single sub-band are then combined, using a hybrid tee to reject high order modes, and the desired signal is routed to a TES island bolometer. Each PA3 pixel has four bolometers for the two linear polarization signals at each frequency. Details about the pixel design can be found in \citet{datta/etal:2014}. We define the detector optical efficiency $\eta$ as the ratio of the power detected by the bolometer's TES to the optical  power incident on the OMT.  Both signal reflections and signal loss can cause $\eta <1$.  We originally projected $\eta \le 0.76$ and $\eta \le 0.66$ for the 97 and 148 GHz bands, respectively.  Table~\ref{tab:params} shows median measured values for the PA3 wafers, indicating the achieved dielectric loss tangents were slightly better than the conservative projections.

For all three arrays, the Nb microstrip lines terminate at a pair of TES islands, where the power is deposited as heat through lossy gold meanders. Each island is connected to the bulk silicon through four SiN legs of length \mbox{61 $\mu$m} and widths that range from \mbox{14 $\mu$m} to \mbox{53 $\mu$m}, depending on the wafer. These legs carry the TES bias/CMB signal lines onto the island and determine the thermal conductivity $G$ to the thermal bath. The superconducting element of the ACTPol TES design is a MoCu ``bilayer" with a $T_{c}$ tuned to approximately \mbox{150 mK} through the choice of the molybdenum and copper geometry and thickness. In the design of the TES, the choice of  $T_{c}$ and $G$ represents a balance between minimizing the thermal noise and increasing the maximum operational optical loading power, $P_{\mathrm{sat}}$.  To increase the stability of the TES through increased heat capacity to slow the detector response, a region of PdAu is coupled to the TES bilayer.   A summary of the detector characteristics and parameter measurements is shown in \mbox{Table \ref{tab:params}}.

\begin{table*}[ht]
\centering
\begin{tabular}{cllllrll}
\toprule
\bf{Array} & \bf{Wafer} & \bf{Type}   & \bf{$T_c$ (mK)} & \bf{$P_\mathrm{sat} $ (pW)} & \bf{G (pW/K)} & \bf{$\eta_{det}$ (\%)}\\
\midrule
\multirow{6}{*}{PA1}&W10 & H & $143 \pm 5$ & $8 \pm 2$ & $226 \pm 30$ & $59 \pm 5$\\
& W09 & H &$143 \pm 5$ & $9 \pm 2$ & $225 \pm 24$ & $61 \pm 8$\\
& W08 & H  & $146 \pm 6$ & $7 \pm 2$ & $173 \pm 33$ & $62 \pm 7$\\
& SH1A & SH & $158 \pm 5$ & $14 \pm 2$ &$326 \pm 25$ & $47 \pm 14$\\
& SH2A & SH & $156 \pm 6$ & $14 \pm 2$ &$327 \pm 24$ & $18 \pm 12$\\
& SH2B & SH & $151 \pm 4$ & $12 \pm 1$ &$303 \pm 19$ & $21 \pm 12$\\
\midrule
\multirow{6}{*}{PA2}&FHC3 & H  &  $186 \pm 5$ &  $10 \pm 1 $  & $ 191 \pm 16 $  &  $60 \pm 3$\\
& FHC1 & H  & $ 199 \pm 5 $ &   $ 16 \pm 2  $  &   $ 276 \pm 48  $ &  $65 \pm 5$ \\
& FH6 & H  &   $ 142 \pm 7 $  &  $14 \pm 4 $ &  $ 469 \pm 111  $ &  $62\pm 2$ \\
& SH3B & SH &  $ 135 \pm 8  $  &  $ 9 \pm 2  $   &  $ 259 \pm 55  $  & $51 \pm 2$ \\
& SH4A & SH &   $ 150 \pm 11  $  &  $ 12 \pm 3  $  &  $ 320 \pm 43  $  & $51 \pm 8$ \\
& SH4B & SH &  $ 129 \pm 8  $ &  $ 15 \pm 5  $  &  $ 436 \pm 138 $ & $51 \pm 2$ \\
\midrule
\multirow{12}{*}{PA3}& \multirow{2}{*}{FH2} & {HA}  & 162 $ \pm  $ 2 & 11 $ \pm $ 1 & 297 $ \pm $ 13 & 83 $ \pm $ 4\\
& & {HB} & 161 $ \pm $ 2 & 12 $ \pm $ 1 & 318 $ \pm $ 17 & 88 $ \pm $ 2\\[0.2 cm]
& \multirow{2}{*}{FH3} & {HA}  & 160 $ \pm $ 2 & 12 $ \pm $ 1 & 314 $ \pm$ 16 & 73 $ \pm $ 6\\
& & {HB}  & 157 $ \pm $ 2 & 12 $ \pm $ 1 & 333 $ \pm$16 & 75 $ \pm $7\\[0.2 cm]
& \multirow{2}{*}{FH4} & {HA}  & 152 $\pm$ 2 & 10 $\pm$ 1 & 293 $\pm$ 15 & 47 $ \pm $4\\
& & {HB}  & 150 $\pm$ 2 & 10 $\pm$ 1 & 302 $\pm$ 15 & 37 $\pm$ 7 \\[0.2 cm]
& \multirow{2}{*}{SH1A} & {SHA}  & 148 $\pm$ 1 & 8  $\pm$ 1 & 240 $\pm$ 10 & 71 $\pm$ 3\\
& & {SHB}  & 146 $\pm$ 2 & 8  $\pm$ 1 &  247 $\pm$ 13  & 79 $\pm$ 2\\[0.2 cm]
& \multirow{2}{*}{SH1B} & {SHA}  & 152 $ \pm $ 2 & 10 $ \pm $ 1 & 249 $ \pm $ 13& 70 $ \pm $ 1\\
& & {SHB}  & 149 $ \pm $ 1 & 9 $ \pm $ 1 & 257 $\pm$ 7& 78 $ \pm $ 2\\[0.2cm]
& \multirow{2}{*}{SH8B} & {SHA}  & 170 $\pm$ 1 &  13 $\pm$ 1 & 314 $\pm$ 11 & 59 $\pm$ 4\\
& & {SHB}  & 169 $\pm$ 1 &  14 $\pm$ 2 & 339 $\pm$ 10 & 71 $\pm$ 4 \\[0.2cm]
\bottomrule
\end{tabular}
\caption{The measured average ACTPol detector parameter values by wafer.  The ``Wafer" column gives unique identifiers  related to the fabrication generation.  The ``Type" column distinguishes hex (H) and semihex (SH) wafers;  for PA3, the additional designator A (B) distinguishes  the 97 GHz (148\,GHz) devices.  We tabulate the mean critical temperatures ($T_c$); the mean bias powers to bring the detectors to 90\% of their normal resistances when the bath temperature is 80~mK in the absence of significant optical loading ($P_{sat}$); the mean thermal conductances from the detector islands to the bath ($G$); and the mean detector efficiencies ($\eta_{det}$).  For each device, we estimate the efficiency as the fraction of the optical power incident on the OMT which is detected by the TES.  Thus the efficiency accounts for losses due to absorption and reflection in the OMT and microstrip on the pixel.  For each quantity, the error listed is the standard deviation of the distribution of devices for that wafer (or wafer/frequency subset for PA3).  These parameters are derived from a combination of measurements in the laboratory and on the telescope.  More detail can be found in \citet{gracethesis}, \citet{ho/etal:2016}, and \citet{pappasthesis}.}
\label{tab:params}
\end{table*}

The response of the ACTPol detectors to a delta function signal diminishes with time due to the electro-thermal time constant, $\tau$, of the TES architecture. The median time constants for the three arrays are \mbox{1.9 ms}, \mbox{1.8 ms}, and \mbox{1.3 ms}.  With a scan speed of \mbox{1.5 deg/s}, these values are equivalent to 3\,dB points at multipoles $\ell$ between roughly 30,000 and 40,000.   The time constants are loading dependent. We find that $f_{\rm 3dB}=1/2\pi\tau$ typically varies by less than 20 Hz/pW.  

Median array sensitivities for 2015 estimated by calibration from Uranus are: {\mbox{$\sim$\,16 $\mu$K\,$\sqrt{s}$} for PA1}, {\mbox{$\sim$\,9.5 $\mu$K\,$\sqrt{s}$} for PA2}, {\mbox{$\sim$\,10 $\mu$K\,$\sqrt{s}$} for PA\,3 (97\,GHz)}, and {\mbox{$\sim$\,14 $\mu$K\,$\sqrt{s}$} for PA3 (148\,GHz)}.  These sensitivity estimates are for PWV/sin(alt)$=$ 1\,mm (because the detectors are photon-noise limited, their sensitivities are dependent on the level of optical loading), and represent the array white noise level, evaluated near 20 Hz. The white noise level of final maps is higher by $\sim$\,60\% than expectations based on projections using these array sensitivities, the reason for which is currently being explored.

\subsection{Readout}

The time-domain multiplexing (TDM) readout scheme employs three stages of SQUID amplification. The multiplexing of the readout has the advantage of reducing the wiring requirements to limit the thermal load on the cold stage.  The current through each TES couples to the input coil of a corresponding SQUID amplifier, called the first-stage SQUID (SQ1).  A set of 32 SQ1s are coupled to one second stage SQUID (SQ2) through a summing coil.  Each SQ2 and its corresponding SQ1s are housed on a multiplexing chip (``MUX11c"). The multiplexing readout is achieved through rapid sequential biasing of each of these 32 SQUIDs.  The final stage of amplification is accomplished by a set of SQUID series arrays (SA).  Twisted pair NbTi cables carry the SQ1, SQ2, and detector lines from the five SA boards housed on the \mbox{1\,K} stage to nine other printed circuit boards (PCBs) on the \mbox{100 mK} stage, which hold the multiplexing and other readout chips.  Each detector is voltage biased onto its transition using an \mbox{$\sim$\,180 $\mu \Omega$} shunt resistor housed on the interface chip, which also contains a \mbox{600 nH} ``Nyquist" inductor designed to band limit the response. 

The detector bias lines are carried from the readout PCBs onto the detector wafers through flexible circuitry consisting of a Kapton base and superconducting aluminum traces connected on both ends by aluminum wire-bonds. Both PA1 and PA2 as well as the hex wafers in PA3 are assembled with two layers of 200 micron pitch flexible circuitry made by Tech-Etch.\footnote[2]{45 Aldrin Rd, Plymouth, MA 02360} For the PA3 semihexes, single-layer, \mbox{100 micron} pitch flex fabricated by the ACT collaboration is used \citep{Pappas:2015}.
The biasing and multiplexing are handled by the Multi-Channel Electronics (MCE) crate (\citealp{battistelli2008}; \citealp{hasselfield:phd}).  From the MCE box, a set of five 100-pin cables carry the bias and readout lines for the SQUIDs and detectors to the SA boards.  The multiplexing and sampling rates are discussed in Section \ref{sec_control}.

The total achieved readout yield for each arrays, where the yield is based on detectors with functional IV curves, is  67\% for PA1, 77\% for PA2, and 83\% for PA3.  The number of detectors used for CMB analysis is typically somewhat less than this (e.g., about 500 detectors for PA1), depending on observing conditions and the number of detectors that can be successfully biased.  The fabrication yield of the detector wafers was high, with many wafers having perfect physical yield of the bolometers.  Most detector loss originated from problematic SQUID biasing and readout lines, each of which corresponds to 32 TESes, and individual detector opens and shorts in the flexible circuitry.  Through improvements in assembly protocols, higher yield was achieved with each successive array.  \mbox{Figure~\ref{fig:array}} shows part of the readout electronics architecture as well as a fully assembled array package.


\section{Cryostat}
\label{sec_cryostat}

\begin{figure*}[t!]
\begin{center}
\includegraphics[width=5in]{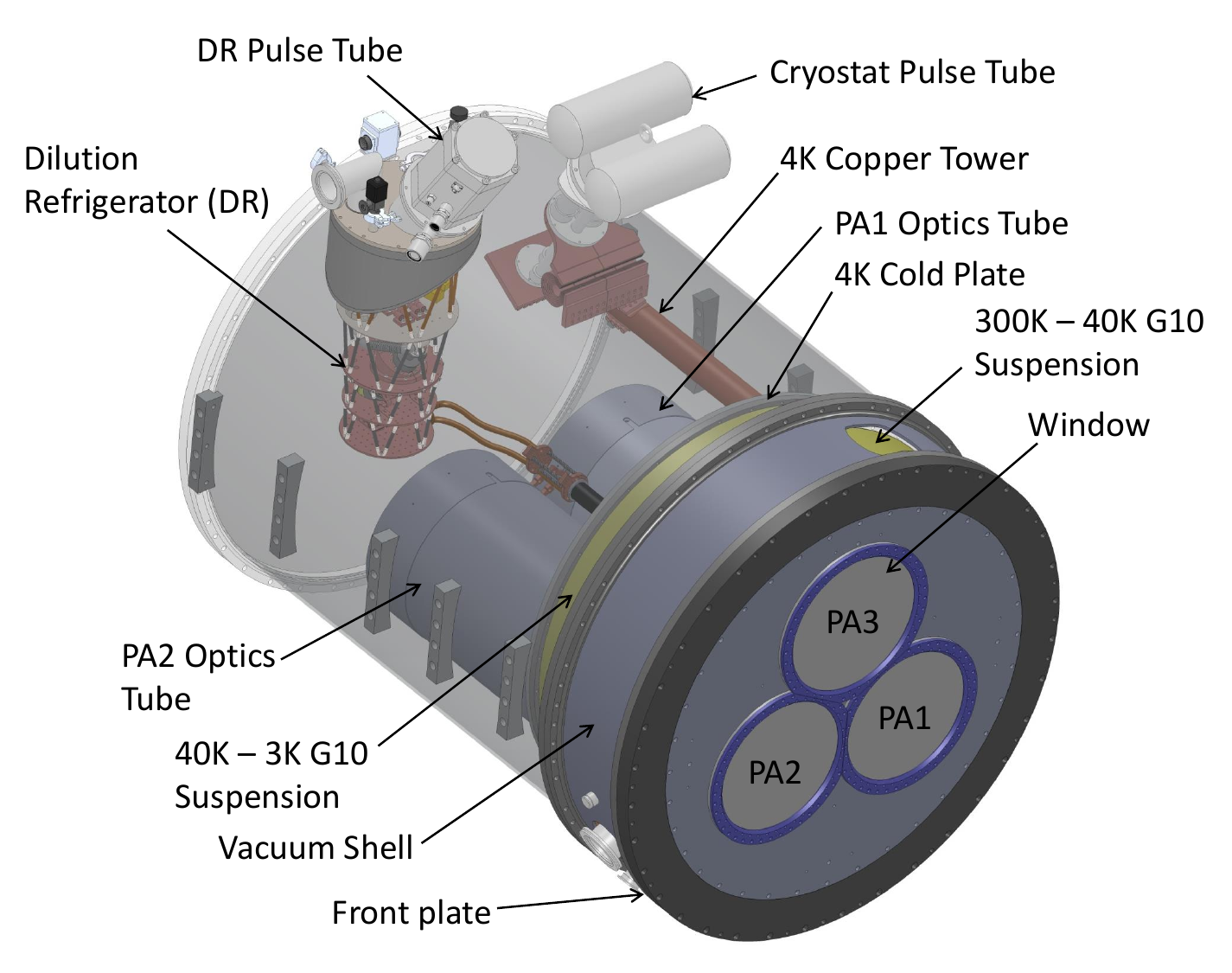}
 \end{center}
\caption{Model of the as-built cryostat.  For scale, the length of the cryostat is 1.5 m.  The PA3 optics tube and most of the radiation shields have been removed for clarity. A combination of flexible copper sheets and copper braid are used to reduce vibrational coupling between the pulse tubes and internal cryostat components.}
 \label{fig:receiver_cutaway}
\end{figure*}

The cryostat is a custom aluminum (primarily alloy 6061) structure fabricated by Precision Cryogenics.\footnote[3]{7804 Rockville Rd, Indianapolis, IN 46214}  Shown in Figure~\ref{fig:receiver_cutaway}, it is a cylinder 1.5\,m in length  and 1.1\,m in diameter.  The size was limited by what could fit inside the existing telescope receiver cabin.     The design was motivated, in part, by the success of the MBAC cryostat (\citealp{swetz:2009}; \citealp{swetz2011}).  The front plate of the shell serves as the optical bench to which all of the cold optics and radiation shields are sequentially mounted.  The 40\,K optics plate, which supports a filter stack (Figure~\ref{fig:cryo_raytrace}), is attached to the cryostat front plate via a G10 fiberglass cylinder.  The 4\,K optics plate is in turn attached to the 40\,K optics plate via a second G10 fiberglass cylinder. The 4\,K optics plate is made out of alloy 1100 aluminum for increased thermal conductivity and acts as the primary support for all of the optics tubes.  Nested 40\,K and 4\,K aluminum radiation shields surround the optics and cryogenics at those temperatures.  Welded to the 40\,K radiation shield are high-purity aluminum strips for improved heat removal from the 40\,K filters.   

\subsection{Cryogenics}

The remote site location makes the use of non-recycled liquid cryogens both difficult and expensive.  Thus, every effort was made to include closed-cycle cooling systems wherever possible.  A single Cryomech\footnote[4]{http://www.cryomech.com.} PT410 pulse tube refrigerator is responsible for cooling the optics and cryostat structures attached to the 40\,K and 4\,K stages (Figure\,\ref{fig:opticstube}).  These include the 40\,K and 4\,K cold plates, radiation shields, 40\,K filters, and the 4\,K components (first lens, baffles, filters) contained within the upper halves of all three optics tubes (Section~\ref{sec_opticstubes}).  To minimize vibrations from the pulse tubes and scan turnarounds, acoustically deadened copper braid and flexible copper sheets are used to attach the PT410 cold stages to the rest of the internal cryostat structures (Figure~\ref{fig:receiver_cutaway}).  The receiver is located near the the telescope azimuthal axis, which also reduces scan-induced vibrations.

All remaining components are cooled below 4\,K using a pulse tube (Cryomech PT407) - backed custom $^3$He\,--\,$^4$He dilution refrigerator (DR; \citealp{shvarts2014}) manufactured by Janis Research Corporation.\footnote[5]{225 Wildwood Ave, Woburn, MA 01801}  The 1\,K stage or ``still" of the DR is responsible for cooling the back end of the optics tubes: the Lyot stop, the second and third lenses, and filters.   The 100\,mK stage or ``mixing chamber" of the DR sets the bath temperature for all three detector arrays and can continuously supply over 100\,$\mu$W
of cooling power at 100 mK, making the DR an excellent choice when compared to more
conventional 100 mK adiabatic demagnetization refrigerators due to the latter's limited cooling power. The DR's 
lower base temperature (thereby improved detector sensitivity) and continuous run-time also out-perform the $^3$He adsorption fridges used in the
MBAC cryostat. 

Cooling the cryostat to base temperature typically takes 14 days with all three sets of optics installed.  This is due in large part to the considerable thermal mass contained within its optical components, focal-plane arrays, and many low-temperature thermal interfaces. The vast majority of the cool-down
($>$ 13 days) is the time required for all components to reach the base temperature of the pulse tube stages.  During this process, the 1\,K and 100\,mK components are connected to the 4\,K stage of the DR pulse tube via a mechanical heat switch. Once the pulse tube base temperatures are reached, the heat switch is disengaged and the $^3$He\,--\,$^4$He mixture is allowed to condense while being circulated within the DR insert to complete the cool-down. An additional 4\,--\,5 hours are needed for the DR mixing chamber to drop below 100~mK once this final step has been initiated. Throughout the cool-down, as well as during normal operations,
the DR is monitored via an ethernet link connected to its gas-handling system computer inside the receiver cabin of the telescope.

The lowest temperature reached by each cryogenic stage depends on a number of operational and environmental factors, including exterior temperature, telescope elevation, scanning motion, and the detector read-out electronics. Table~\ref{tab:cryoload}  lists the measured base temperatures and estimated thermal loading for each cryogenic stage during optimal cooling conditions -- note that the temperature of the coldest stage (the DR mixing chamber) may be up to 13\,mK warmer during typical observing operations (telescope in motion and detector read-out powered on). Since a lower bath temperature results in higher detector saturation powers
(and thus a larger dynamic range over atmospheric loading conditions (see Section~\ref{sec_arrays}), gold-plated high-purity (99.999\%) annealed copper links were used to make thermal connections between the DR and the focal-plane arrays to minimize temperature gradients. During observations, the typical array temperature with all three sets of optics installed ranged from 100 to 115 mK.

\begin{table}[tbh]
\label{tab:cryo_load}
\centering
\begin{tabular}{ccc}
\midrule

Cryogenic Stage                    & Temperature & Load  \\
\midrule

PT410 1st Stage               & 34.1 $\pm$ 0.7~K       & 9.7 $\pm$ 2.2~W \\

PT410 2nd Stage               & 3.65 $\pm$ 0.05~K      & 620 $\pm$ 30~mW \\

DR Still                        & 644 $\pm$ 4~mK     & 1.09 $\pm$ 0.10~mW \\

DR Mixing Chamber         & 82.4 $\pm$ 0.5~mK     & 53.4 $\pm$ 1.4~$\mu$W \\

\midrule

\end{tabular}
\caption[Temperatures and loading of the different temperature stages]{Measured base temperatures and thermal loads of the receiver cryogenic stages with three sets of optics during optimal cooling conditions (detector electronics powered down and telescope at rest).}
 \label{tab:cryoload}
\end{table}

\subsection{Optical Support Structures}
\label{sec_opticstubes}

\begin{figure*}[!t] 
\begin{center}
\includegraphics[width=6in]{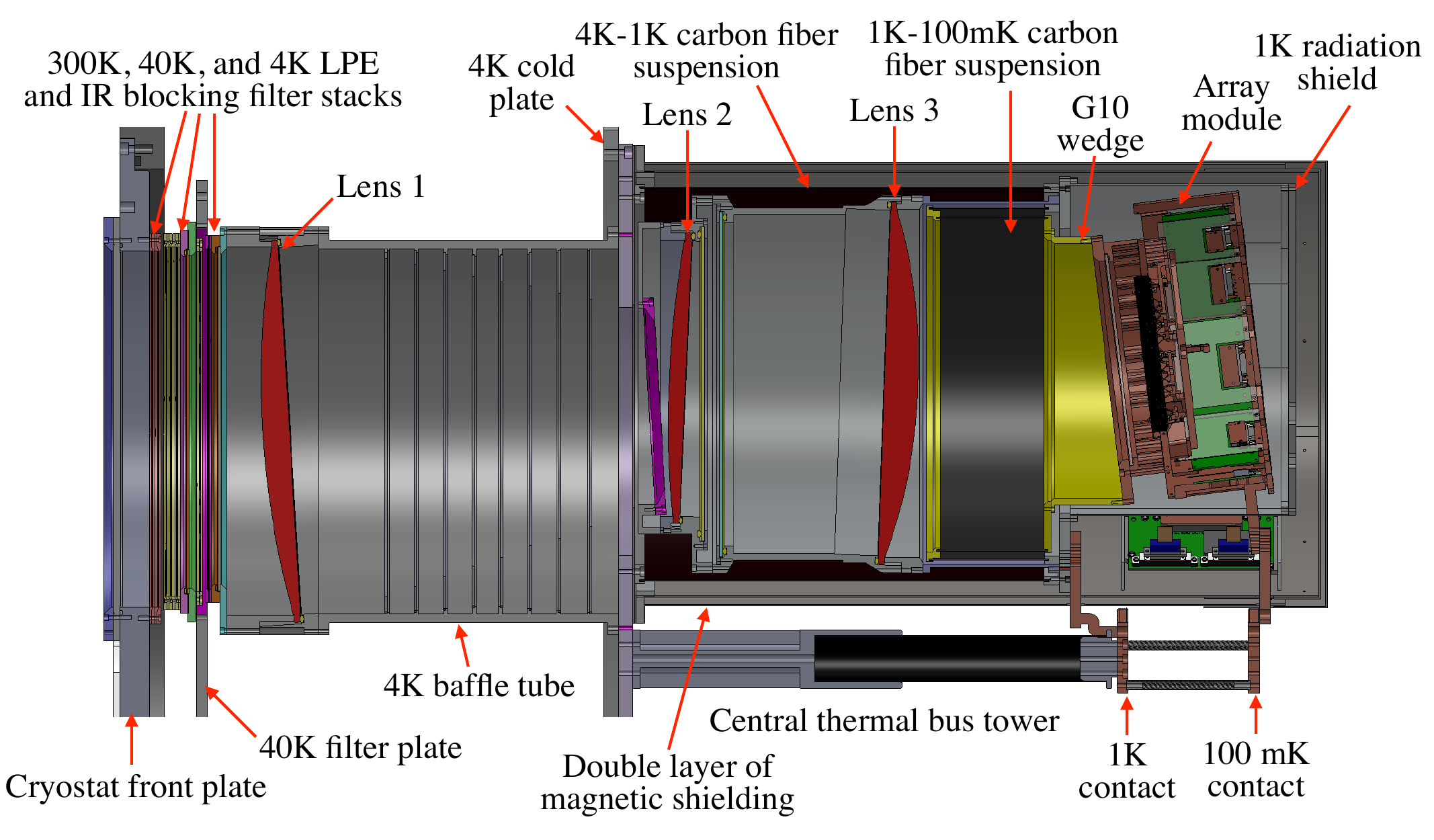}
 \end{center}
\caption{Cutaway view of the PA3 optics tube showing the internal optics, mechanical structures, magnetic shielding, and cold straps.}
 \label{fig:opticstube}
\end{figure*}

Each of the three sets of optics is contained in a cylindrical optics tube. To minimize weight, aluminum was generally used to mount optical elements between  300\,K and 1K.  For 100\,mK assemblies, where conductivity is critical and where aluminum is superconducting (resulting in reduced thermal conductivity), oxygen-free high-conductivity copper (OFHC) was generally used. Another reason for using mostly copper below $\approx$ 1\,K is potential problems with magnetic flux expelled by superconducting aluminum alloys, which is problematic for the SQUIDs (Section~\ref{sec_magshielding}).

The three optics tube assemblies are of similar design, yet self-contained to allow each one to be deployed individually. A cross-section of the PA3 optics tube is shown in Figure~\ref{fig:opticstube}.  All lenses and LPE filters are clamped with a spiral beryllium copper spring around the perimeter of one side of each element to accommodate differential thermal contraction and to provide a uniform clamping force against the brittle silicon. The spring is a commercial\footnote[6]{Internet URL: http://www.spira-emi.com/} product designed primarily for shielding from electromagnetic inteference. Most of the cylindrical structures comprising each optics tube are mounted perpendicular to the cold plates. Individual lens and array tilts were achieved through custom angled mounts that hold each component at the proper angle with respect to each optics tube.  

The first lens, nominally at 4\,K, is supported from the 4\,K cold plate via an aluminum tube, inside of which are machined steps that precisely locate baffles.  The baffles are blackened with a mixture of 2850 FT Stycast, loaded with 5\,-\,7\,\% carbon lampblack by weight, so that the minimum thickness exceeds 1\,mm and is textured.  The Lyot stop, second and third lenses, and surrounding low-pass filters are all maintained near 1\,K.  To reduce the load on the DR, these structures are supported from the 4\,K cold plate using a custom-made carbon fiber tube.  Attached to this 1\,K framework is a second, re-entrant carbon fiber suspension that is 0.6\,mm thick for thermally isolating the 100\,mK array package from the 1\,K components.  

\subsection{Magnetic Shielding}
\label{sec_magshielding}

The SQUID multiplexers and amplifiers, as well as the TESes themselves, are sensitive to changing magnetic fields and therefore require magnetic shielding from both Earth's field and AC fields associated with the telescope motion. To complement shielding around the SA modules and at the PCBs, additional magnetic shielding was provided for the collection of SQUIDS for each array by enclosing the lower half of each optics tube (Figure~\ref{fig:opticstube}) with  Amumetal 4\,K\footnote[7]{A trademark of Vacuumschmelze GmbH in Hanau, Germany.
Local Distributor: Amuneal Manufacturing Corporation, 4737 Darrah St.,
Philadelphia, PA 19124, USA, info@amuneal.com, (800)-755-9843.} (A4K).  This is a proprietary material with a high nickel concentration and prepared with a heat treatment process to give it a high magnetic permeability at cryogenic temperatures. To achieve the maximum attenuation, the thickest available A4K (1.5\,mm) was used.  Each optics tube has two layers of shielding separated by approximately 6\,mm since, given sufficient spacing, using additional layers approaches the limit of multiplicative increases in the field attenuation.

\subsection{Receiver Alignment to Telescope}
\label{sec_alignment}

The receiver is supported and aligned by an aluminum frame that was designed to interface with the receiver cabin structure (Figure~\ref{fig:telescoperaytrace}).  Plain (sliding) phosphor bronze bearings support the receiver and allow it to be translated in three directions and rotated in two directions.    The positions of the telescope optics and receiver are measured using a VSTARS photogrammetry system from Geodetic Systems Inc. (GSI).\footnote[8]{1511 Riverview Dr, Melbourne, FL 32901}  Retroreflective targets are placed at the location of the alignment actuators on each panel of the primary and secondary mirrors as well as at fiducial points on the front surface of the cryostat.  Several auxiliary targets are placed around the perimeter of each mirror and on the inner ground screen.  The target 3D positions are solved for using GSI propriety software.  After post-processing with in-house software and actuator adjustment, the mirror surfaces are aligned to better than 27 $\mu$m RMS during nighttime observing and the reciever is positioned to within 1\,mm of its optimal location.


\section{Data Acquisition}
\label{sec_control}

\subsection{Overview}
Both science data and housekeeping data are recorded.  The science data consist of detector output from the three arrays, with data from each written to a separate local acquisition computer. A housekeeping computer logs data on ambient and cryogenic temperatures, telescope encoders and motors, etc.  Due to the dusty environment and low atmospheric pressure, all of the hard drives at the telescope site are solid state. Data from each acquisition machine are transmitted via radio link to a RAID server at our base near San Pedro de Atacama, where they are merged together into a single data product. The data are ultimately archived in North America.

\subsection{2013 and 2014 Seasons}
\label{ssec:readout_2013-14}

Data acquisition for ACTPol in the 2013 and 2014 seasons was largely identical 
to that of MBAC, so here we only summarize the acquisition and flow of data, 
and refer the reader to Swetz et al. (2011) for details. In the 2015 season some
systems were modified, and we outline those in Section~\ref{ssec:readout_2015}

The MCE bolometer readout has a raw sampling rate of 50 MHz.  Using TDM, with a row-switching rate of 500 kHz and 33 row selects, an entire array of detectors is sampled at 15.15 kHz.  The signal is then anti-alias filtered with a four-pole, low-pass, Butterworth design with a 120\,Hz cutoff.  The data are downsampled to 399 Hz before transmission by optical fiber to the data acquisition computer for recording to disk. Each detector array has its own MCE, read out by and recorded on its 
own computer. Synchronization of the MCEs is achieved by means of a `sync box' that provides a common clock for the 15.15 kHz sampling, as well as an integer
`sync counter' that increments at the 399 Hz recording rate. The data for each MCE are tagged with the sync counter so that they can be merged downstream.

The same sync counter is used to synchronize the readout of the azimuth and elevation encoders with the bolometer channels. Our housekeeping data acquisition computer contains a custom-built PCI card with an onboard FPGA that receives the sync counter from the sync box via an RS485 connection (clock plus NRZ). Reception of a new sync word generates an interrupt on the PCI bus, which alerts a device driver to poll the azimuth and elevation encoders using a Heidenhain IK220 PCI card in the same computer. The driver then bundles the sync counter with the encoder readings and the absolute time---the latter is accurate to $<$ 1 ms due to a Meinberg GPS-169 PCI card disciplining the computer's system clock.

The ACT master control program (AMCP) is responsible for reading and overall control of all housekeeping. The sync box-disciplined output of the device driver described above provides AMCP with the master clock. Cryogenic temperatures are read out at 10.6 Hz using three National Instruments\footnote[9]{http://www.ni.com/en-us.html} USB-6218 DAQs; although this readout is asynchronous with the the sync box, we have measured that they are normally aligned to $<$ 100 ms (1 sample at 10 Hz), which is sufficient for our needs. Ambient temperatures of the telescope structure are read with a Sensoray\footnote[10]{7313 SW Tech Center Dr, Portland, OR 97223} 2600 at 50 Hz, also asynchronously. The state of the motion control robot as well as its own housekeeping (such as motor temperatures and currents) are also acquired by AMCP. At the same time, ACMP dispatches commands to the motion robot that it receives from human operators
or the automatic scheduler.

The four datasets from our acquisition computers\,-\,three MCE computers and one housekeeping computer\,-\,are copied via radio link to our base near San Pedro de Atacama, where they are merged together by means of of the sync counter. The result is a single archive of data files, all sampled at 399 Hz (or integer factors thereof) and, for the encoder and bolometer data, acquired simultaneously to 5\,$\mu$s precision. Having this merged data product means that mapmaking software need not search for disparate files and can work with fully-synchronized data.

The presence of new data files is registered in a MySQL database, and a series of software daemons that communicate with this database copies files to portable hard disks for transport to Santiago and North America. The software tracks all file copies and ensures that redundant copies exist before space is freed on site computers. This is true for merged files as well as raw, unmerged files which we keep for full redundancy. One can also request that files be copied over the internet rather than by transport drives; our internet bandwidth is too small, however, to forego the use of transport drives.

\subsection{Readout Systems for 2015}
\label{ssec:readout_2015}

Modifications to our housekeeping readout were implemented before the 2015 season of observations to prepare for the installation of ambient-temperature half-waveplates (HWP), with the goal of significantly reducing atmosheric and instrument 1/f noise (e.g., \citealp{Kusaka2014}).  

The HWP encoders will be read out by a DNA-PPC8 system, which consists of a CPU interfaced to a lower data acquisition layer. The latter is customizable by installing various boards provided by the same vendor\footnote[11]{http://www.ueidaq.com/dna-ppc8.html} in the unit's six available slots. Reading our high-speed serial sync data would be challenging with this device, so we have built a new module for the sync box that translates the serial RS-485 sync counter to a parallel signal latching at the 399 Hz data acquisition rate. This parallel sync counter is read by a DNA-DIO-403 card in the DNA-PPC8 and any data acquired by the unit can be tagged with the sync counter.

We took advantage of our ability to interface the DNA-PPC8 with our sync box to update our azimuth and encoder readout as well, as the interrupt-generating scheme described above (Section~\ref{ssec:readout_2013-14}) was quite demanding on our housekeeping computer, causing various complications. The new readout uses a Heidenhain EIB 741 unit, externally triggered at 399 Hz by the DNA-PPC8 over RS485 whenever a new sync counter is received. The azimuth and elevation encoder positions are then read from the EIB 741 unit over ethernet by AMCP. A trigger counter provided by the EIB 741 along with the encoder readings makes it straightforward to associate the latter with the appropriate sync counter, thereby ensuring synchronization of all the encoders---azimuth, elevation, and HWP---with our bolometer readings.


\section{On-Sky Beams and Polarization}
\label{sec_beamsandpol}

\subsection{Beams}
\label{sec_beams}

\begin{figure*}[t!]
\begin{center}
\includegraphics[width=5in]{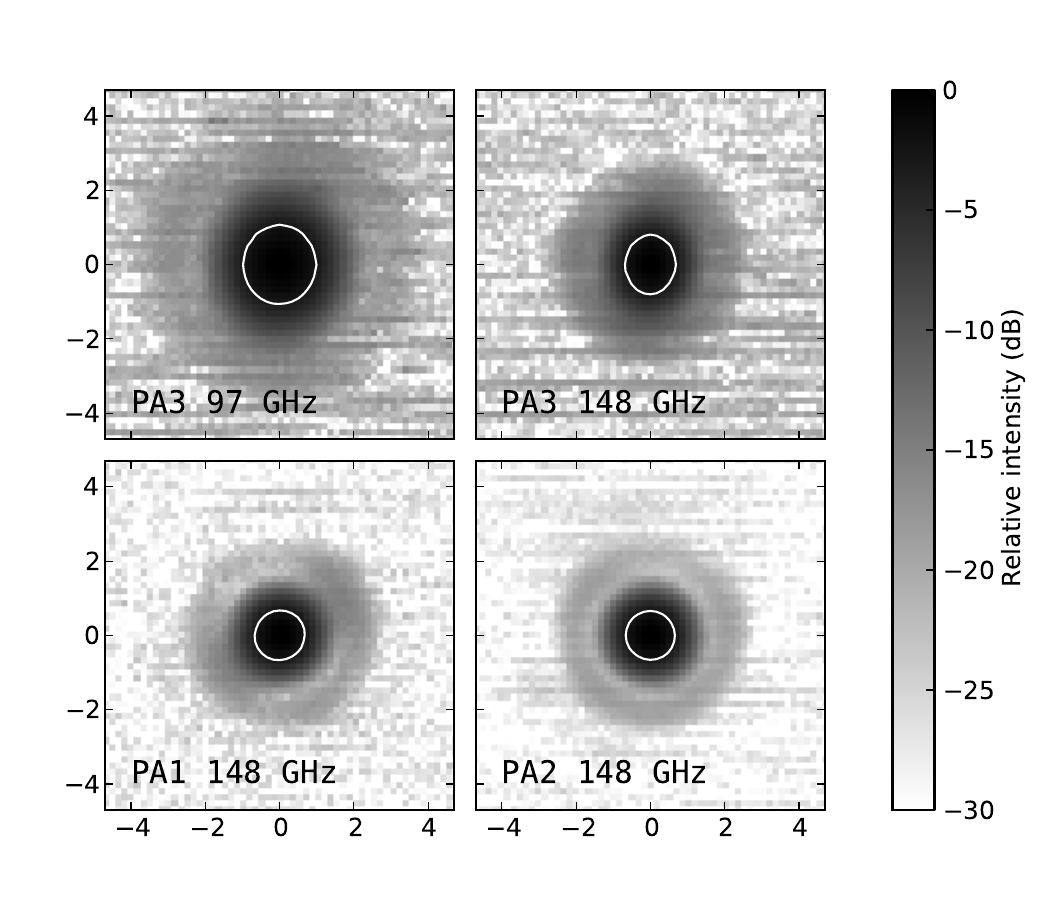}
  \caption{Beam maps for the ACTPol arrays, showing the
    response to a point source in a coordinate system such that North
    is parallel to increasing altitude and West is parallel to
    increasing azimuth.  These maps include the data from all
    responsive detectors, averaged over 81, 86, 39, and 37
    observations for the PA1, PA2, PA3/148, and PA3/97 arrays,
    respectively.  The white contour lines denote where the response
    falls to half of its peak value.}
    \label{fig:beams}
\end{center}
\end{figure*}

\begin{table*}[ht]            
\centering
\begin{tabular}{lcccc}
\midrule
Array -- Band & Solid Angle  & \multicolumn{3}{c}{Elliptical FWHM model} \\
              &              & Major axis & Minor axis & Major angle \\
              & (arcmin$^2$) & (arcmin) & (arcmin) & ($^\circ$) \\
\midrule
PA1 -- 148 GHz & $2.27 \pm 0.05$ & 1.37 & 1.32 & -49\\
PA2 -- 148 GHz & $2.12 \pm 0.03$ & 1.33 & 1.31 &  56\\
PA3 -- 148 GHz & $3.12 \pm 0.13$ & 1.58 & 1.33 &  -3\\
PA3 -- 97 GHz  & $6.62 \pm 0.22$ & 2.13 & 2.00 &  -2\\
\midrule
\label{tab:beams}
\end{tabular}
\caption{Beam parameters, as derived from the maps in
  Figure~\ref{fig:beams}. The elliptical FWHM model parameters are fit
  to points in the map which show a response within 10\% of the
  half-maximum level. FWHM major angle is measured from the North,
  increasing towards the East.} 
\end{table*}

The telescope beams are characterized with observations of planets.
Saturn is used to align and focus the secondary mirror at the start of
each season, but observations of Uranus are ultimately used for beam
characterization because its brightness is low enough to not saturate the detectors.

A single planet observation is achieved by scanning the telescope back
and forth in azimuth, at fixed elevation, while the planet rises or
sets through an array's field of view.  The resulting time-ordered
data are reduced to a single map of the celestial sky, centered on the
planet, for each detector array and frequency band.  This mapping
process relies on knowing the relative pointings of the detectors on
the sky, and on an accurate calibration of each detector's response to
the source.

The detector positions are determined from Saturn and Uranus
observations by fitting a model for the source to the time-stream
data directly.  A single average pointing template is computed for
each season and array, and used for all subsequent planet mapping.
The relative positions of the detectors are constrained at roughly the
arcsecond level, and thus the pointing template uncertainty contributes
negligibly to beam degradation.

The detector calibrations are determined independently for each mapped
observation by comparing the amplitude of each detector's response to
the common mode from atmospheric emission.

Each planet map provides a measure of the telescope's
``instantaneous'' beam, averaged over all responsive detectors in each
array and band.  The average instantaneous beams are shown in
Figure\,\ref{fig:beams}, and some basic properties are provided in
Table\,\ref{tab:beams}.  The effective point spread function of the
telescope in the CMB survey maps differs from the instantaneous beam
primarily due to the impact of pointing variance, which acts as a
low-pass spatial filter, and due to the way in which sky rotation
symmetrizes the beam by stacking up observations at a variety of
parallactic angles.  For the interpretation of CMB survey maps,
the beam's harmonic transform and its covariance are determined
following the procedure described in \citet{Hasselfield2013Beams}.

\subsection{Polarization Angles}
\label{sec_pol}

\begin{figure*}
\begin{center}
\includegraphics[width=6in]{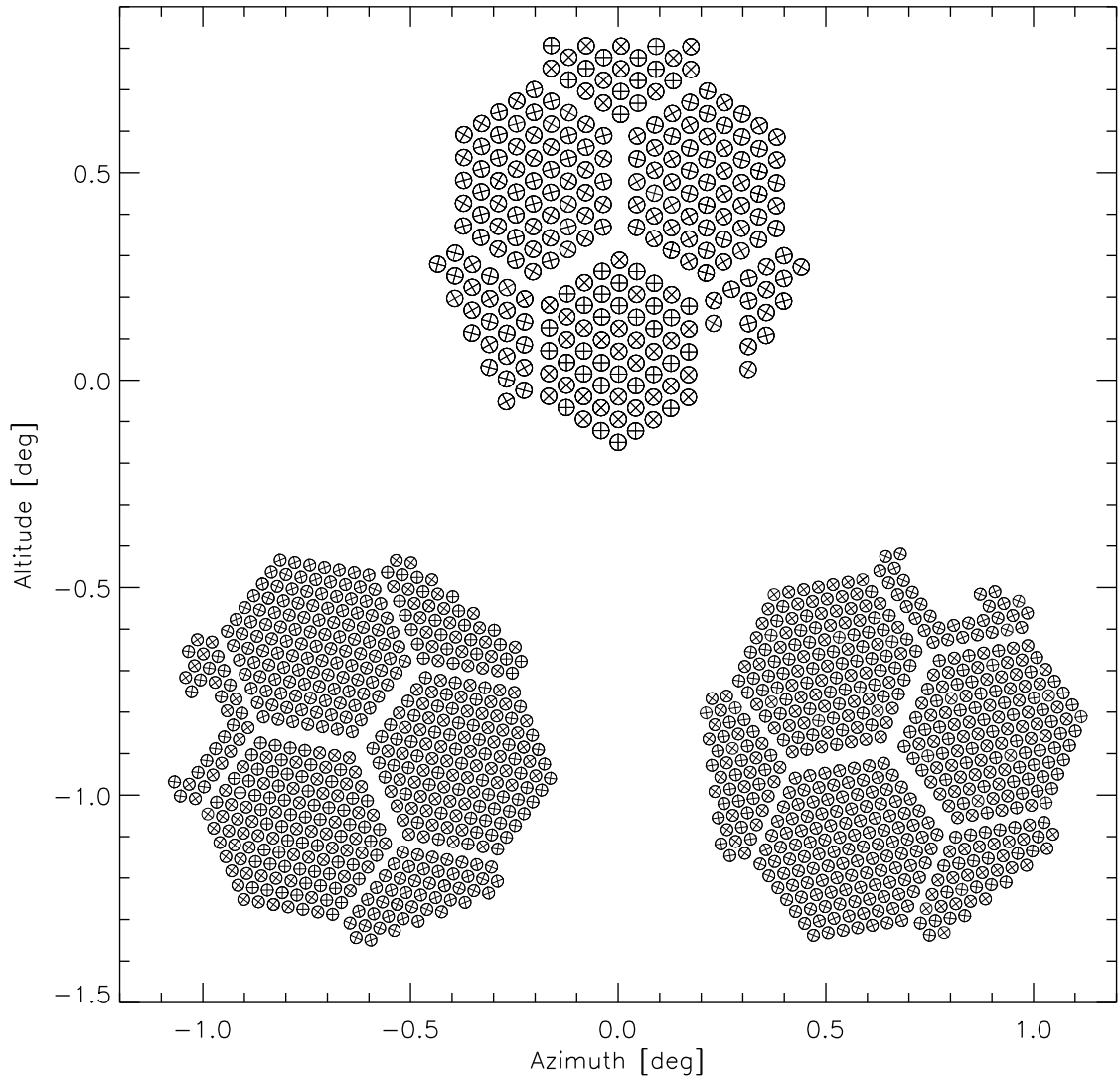}
  \caption{Plot of relative detector positions and polarization angles as seen on the sky.  PA1 is in the lower right, PA2 the lower left, and PA3 on top.  Each hex/semihex wafer has OMT pairs in two different orientations, which results in six different detector polarization angles in each array.    There is a slightly higher number of physical detectors than there are available readout lines for, resulting in the region of seemingly ``missing" detectors in each array.}
    \label{fig:polangles}
\end{center}
\end{figure*}

The ability to separate polarized intensity into E and B components depends upon how accurately detector polarization angles are known.    For each hex and semihex wafer, a feed horn couples to an OMT pair having one of two possible orientations, which is set by lithography during fabrication.  When this is combined with both the orientation of the semi-hex and hex wafers within an individual array as well as the orientation of each array as a whole into the cryostat, a roughly equal distribution of detector angles, covering every 15 degrees,  is produced.   Figure~\ref{fig:polangles} shows these resulting detector polarization angles as seen on the sky. 

Apart from the physical OMT orientations, however, there is also a polarization rotation introduced by the ACTPol optical chain
as seen on the sky.  The optics-induced rotations, as well as the detailed mapping from the focal plane position to projected angle on the sky, are determined using a polarization-sensitive ray trace through a model of the telescope in the optical design software CODE\,V.\footnote[12]{http://www.synopsys.com} Given a point on the sky, the software calculates the polarization state across the entrance pupil diameter. This polarization
state, averaged across the pupil and propagated to the focal plane, produces the polarization output for a single point on the focal plane. Given a perfectly polarized input state of known angle, the difference between the output polarization state and the input state results in the polarization rotation. The resulting polarization rotation output is produced at twenty five points on the focal plane for each of the three ACTPol arrays. 
This rotation on the focal plane is then fit to a simple 2D quadratic model, which is in turn used to produce the polarization angle rotation at the location of each feed horn in all three arrays.  The optics-induced rotations have a range of approximately 3$^\circ$.  While this is too small to be seen in Figure\,\ref{fig:polangles}, it is too large to be ignored in the anaylsis.   

Thus far the measured polarization angles from the Crab Nebula and minimizing the EB spectrum (\citealt{Naess2014}) are consistent with the calculated polarization angles based on the optics design and detector layout.  Further analyses will presented in future papers, e.g. \cite{koopman/etal:2016}.

\section{Conclusion}

We have presented the design and performance of the ACTPol instrument. The new receiver provides a number of improvements to the ACT experiment: polarization-sensitivity, continuous 100\,mK cooling, meta material AR coatings, and multichroic array technology.  ACTPol acquired three seasons of data from 2013 - 2016.  First analyses of Seasons 1 and 2 are presented in \citet{Naess2014} and \citet{Louis2016}.


\acknowledgements

We thank AMEC/Dynamic Structures/Empire for their work on the initial telescope construction and Kuka Robotics for their continued support on the motion control system. Vladimir Shvarts and the rest of the Janis Research Corporation were essential for the successful integration of the dilution refrigerator into the receiver.  We are very grateful for Bill Dix, Glen Atkinson, and the rest of the Princeton Physics Department Machine Shop, and Harold Borders and Jeff Hancock at the University of Pennsylvania Physics
Department Machine Shop. Jesse Treu served as a management consultant for the latter half of the project. 

ACT is on the Chajnantor Science preserve which was made possible by the Chilean Comisi\'on Nacional de Investigaci\'on Cient\'ifica y Tecnol\'ogica. We are grateful for the assistance we received at various times from the ABS, ALMA, APEX, ASTE, and POLARBEAR groups. The PWV data come from the public APEX weather site.  Field operations were based at the Don Esteban (operated by Astro-Norte) and RadioSky facilities.  This work was supported by the U.S. National Science Foundation through awards AST-0408698 and AST-0965625.  Funding was also provided by Princeton University, the University of Pennsylvania, Cornell University, the Wilkinson Fund and the Mishrahi Gift.  The development of multichroic detectors and lenses was supported by NASA grants NNX13AE56G and NNX14AB58G.  CM acknowledges support from NASA grant NNX12AM32H. BS, BK, CM, EG, KC, JW, and SMS received funding from NASA Space Technology Research Fellowships.  RD thanks CONICYT for grants FONDECYT 1141113 and Anillo ACT-1417.

This paper includes contributions from a United States government agency and is not subject to copyright.

\bibliographystyle{apj}
\bibliography{actpolbib}

\end{document}